\documentclass{aa}
\usepackage{graphicx,supertabular}
\usepackage{txfonts}
\usepackage{natbib}
\begin{document}
   \title{Towards a library of synthetic galaxy spectra and preliminary results of classification and parametrization of unresolved galaxies for Gaia}

   \author{P. Tsalmantza\inst{1}
          \and M. Kontizas\inst{1}
          \and C. A. L. Bailer-Jones\inst{2}
          \and B. Rocca-Volmerange\inst{3,4}
          \and R. Korakitis\inst{5}
          \and E. Kontizas\inst{6}
          \and E. Livanou\inst{1}
          \and A. Dapergolas\inst{6}
          \and I. Bellas-Velidis\inst{6}
          \and A. Vallenari\inst{7}
          \and M. Fioc\inst{3,8}}

    \offprints{P. Tsalmantza\\
    \email{vivitsal@phys.uoa.gr}}

    \institute{Department of Astrophysics Astronomy \& Mechanics, Faculty
               of Physics, University of Athens, GR-15783 Athens, Greece                 
         \and
              Max-Planck-Institut f\"ur Astronomie, K\"onigstuhl 17, 69117 Heidelberg, Germany
         \and
              Institut d'Astrophysique de Paris, 98bis Bd Arago, 75014 Paris, France
         \and
              Universit\'e de Paris-Sud XI, I.A.S., 91405 Orsay Cedex, France              
         \and
              Dionysos Satellite Observatory, National Technical University of Athens, 15780 Athens, Greece
         \and
              IAA, National Observatory of Athens, P.O. Box 20048, GR-118 10 Athens, Greece
         \and 
              INAF, Padova Observatory, Vicolo dell'Osservatorio 5, 35122 Padova, Italy
         \and
              Universit\'e Pierre et Marie Curie, 4 place Jussieu, 75005 Paris, France}

\date{Received date / accepted}

% \abstract{}{}{}{}{}
% 5 {} token are mandatory

  \abstract
  % context heading (optional)
   {} %leave it empty if necessary
  % aims heading (mandatory)
   {The Gaia astrometric survey mission will, as a consequence of its scanning
     law, obtain low resolution optical (330--1000\,nm) spectrophotometry of
     several million unresolved galaxies brighter than V=22. We present the
     first steps in a project to design and implement a classification system
     for these data. The goal is both to determine morphological classes and
     to estimate intrinsic astrophysical parameters via synthetic templates.
     Here we describe (1) a new library of synthetic galaxy spectra, and (2)
     first results of classification and parametrization experiments using
     simulated Gaia spectrophotometry of this library.}
  % methods heading (mandatory)
   {We have created a large grid of synthetic galaxy spectra using the
     P\'EGASE.2 code, which is based on galaxy evolution models that take into
     account metallicity evolution, extinction correction, emission lines
     (with stellar spectra based on the BaSeL library). Our classification and
     regression models are Support Vector Machines (SVMs), which are kernel-based
     nonlinear estimators.}
  % results heading (mandatory)
   {We produce a basic library of about 3600 zero redshift galaxy spectra covering
     the main Hubble types over wavelength range 250 to 1050\,nm at a sampling
     of 1\,nm or less. It is computed on a regular grid of four key
     astrophysical parameters for each type and for intermediate random values
     of the same parameters. An extended library reproduces this at a series
     of redshifts. Initial results from the SVM classifiers and parametrizers are
     promising, indicating that Hubble types can be reliably predicted and
     several parameters estimated with low bias and variance. Comparing the colours 
     of our synthetic library with Sloan Digital Sky Survey (SDSS) spectra we
     find good agreement over the full range of Hubble types and parameters.}
  % conclusions heading (optional), leave it empty if necessary
   {}
   

\maketitle

\section{Introduction}
Large surveys of galaxies provide information on their global spatial
distribution and the physical properties of individual galaxies. Such a survey
will be obtained for the whole sky by the ESA mission, Gaia, from 2011--2016.
During its five year mission Gaia will observe several million unresolved
galaxies all over the whole sky. Although the survey's main goal is the
stellar content and the structure of our galaxy, there remains a lot of
important science to be extracted from the galactic component.
                                            
There currently exist several surveys of galaxies, but even SDSS -- one of the
most extended galaxy photometric and spectroscopic surveys in the the optical
and near IR (about at the spectral range of Gaia) -- covers only a fifth of
the sky. Gaia extends this in several ways: i) It will be able to detect about
10$^7$ unresolved galaxies down to G=20 (V=20--22); ii) Gaia will be the first
homogeneous survey of galaxies covering the whole sky since photographic ones
(UK, ESO, Palomar Schmidt surveys, 3500 to 6500\,\AA) of 30 years ago; iii) The
spectrophotometry covers a larger spectral range (3300 to 10\,000\,\AA\ 
sampled in about 100 bins) than earlier surveys; iv) Gaia observes each source
an average of 80 times over the mission. With this we can investigate many
different types of galaxy, QSO and AGN variability; v) The sample will have a
well-defined selection function, important for estimating the galaxy density
in the local universe.

Our long-term objective is to classify and to determine the astrophysical
parameters of all unresolved galaxies which Gaia will observe. In order to
proceed with this we first need to acquire or build an appropriate library of
galaxy spectra. This library must show sufficient variation in those
intrinsic astrophysical parameters (APs) to which the Gaia observations will
be sensitive. To determine APs on a homogeneous system we ultimately need to
build or calibrate our classifiers using synthesis models and synthetic
spectra. Existing observed or synthetic libraries are too small or don't
cover the required wavelength range. For this reason we set on in this paper
to start building a new library.

We use the galaxy evolution model P\'EGASE (Projet d' Etude des Galaxies
par Synthese Evolutive) \citep{fioc2,fioc5}, to synthesize galaxy spectra. The
P\'EGASE.2 code\footnote{http://www2.iap.fr/users/fioc/PEGASE.html} is aimed
principally at modelling the spectral evolution of galaxies by types: the
active and passive evolution of stellar populations as well as interstellar gas
and dust are coherently evolved in time. No galaxy number density evolution is
considered, although the results of our models are compatible with occasional
rare galaxy merging. The code is based on the stellar evolutionary tracks from
the Padova group, extended to the thermally pulsating asymptotic giant branch
(AGB) and post-AGB phases \citep{groenewegen}. These tracks cover all the
masses, metalicities and phases of interest for galaxy spectral synthesis.
P\'EGASE.2 uses the BaSeL 2.2 library of stellar spectra and can synthesize low
resolution (R=200) ultraviolet to near-infrared spectra of Hubble sequence
galaxies, as well as of starbursts. For a given evolutionary scenario
(typically characterized by a star formation law, an initial mass function and,
possibly, infall or galactic winds), the code consistently gives the spectral
energy distribution (SED) and computes the star formation rate and the
metallicity at any time. The nebular component (continuum and lines) due to HII
regions is calculated and added to the stellar component. Depending on the
geometry of the galaxy (disk or spheroidal), the attenuation of the spectrum by
dust is then computed using a radiative transfer code (which takes account of
the scattering).

By accepting a star formation rate proportional to mass of the gas, the IMF of
\citet{rana} and the presence of infall and galactic winds, eight synthetic
spectra corresponding to different typical types of Hubble sequence galaxies
(E, S0, Sa, Sb, Sbc, Sc, Sd and Im) have already been produced using P\'EGASE.2
\citep{fioc3,fioc1,le2}. For each type, the values of the parameter set have
been fitted to the observed spectral energy distribution (SED) of nearby (z=0)
galaxies. For illustration a comparison with data is shown in \citet{fioc6}. At
higher redshifts, the evolution scenarios have been tested against most
existing faint galaxy samples, including the deepest surveys \citep[B=29 Hubble
Deep Field-N,][]{williams}. One unique model of galaxy fractions by type
simultaneously predicts the multi-wavelength (UV to near-IR) galaxy counts,
dominated by young stellar populations in the UV and old evolved galaxies in
the near-IR respectively.  The faint blue galaxy population, in excess in the
far-UV, has also been analysed \citep{fioc4}. An episodic star formation rate
of low level is proposed to fit the far-UV counts \citep{FOCA2000,buat}. In the
near-IR, the evolution scenario of elliptical galaxies predicts the puzzling
$K$-$z$ relation of radio galaxy hosts between z=0 and z=4. \citet{rocca} use
P\'EGASE.2 scenarios to interpret the galaxy distribution in the K-band Hubble
diagram. The same models are used to interpret the mid-IR galaxy counts \citep{rocca07},
although here a supplementary ultra-luminous infrared galaxy population is
required. Finally, the robustness of our evolution scenarios is confirmed by
the significant predictions of photometric redshifts as compared to
spectroscopic redshifts of HDF-N sample \citep{le2}. Using a much larger sample
from the SDSS, we make an additional comparison. This is the subject of the
second section of this paper, made using simulated photometry and colour-colour
diagrams.  In section 3 we describe the production of our library based on
these eight typical synthetic spectra of galaxies and in section 4 we explain
how these are used to simulate Gaia data. In section 5 we present our
classification and parametrization models and give preliminary results on their
performance.  A brief discussion follows in section 6.

\section{P\'EGASE synthetic spectra and comparison with the SDSS spectra}

In order to determine the parameter ranges over which we should generate the
library, we first make a comparison of colours synthesized from the eight
typical P\'EGASE spectra with SDSS data. To avoid small discrepancies that
occur between synthesized and published SDSS
photometry\footnote{http://www.sdss.org/dr4/products/spectra/spectrophotometry.html}
and to treat both types of spectral data in the same way, we decided to
synthesize SDSS photometry from the SDSS spectra in the same way as we do with
the synthetic spectra (and using the same ``calib'' and ``colors'' programs in
the P\'EGASE.2 code for both). For this we use the whole set of spectroscopic
data for the 565\,715 galaxies that are available in data release 4 (DR4) of
SDSS. The properties of the SDSS filters are given in Table~\ref{t1}.

\begin{table}
 \centering
 \caption {Characteristics of the five SDSS filters}                                      
 \begin{tabular}{c c c c c}          
 \hline\hline                        
  Name & Average    & Starting   & Ending     & magnitude \\
       & wavelength & wavelength & wavelength & limit in  \\
       & ($\AA$)    & ($\AA$)    & ($\AA$)    & survey    \\
 \hline
u      & 3551       & 2980       & 4130       & 22.0      \\
g      & 4686       & 3630       & 5830       & 22.2      \\
r      & 6165       & 5380       & 7230       & 22.2      \\
i      & 7481       & 6430       & 8630       & 21.3      \\
z      & 8931       & 7730       & 11230      & 20.5      \\
\hline
\end{tabular}
\label{t1}
\end{table}

\begin{figure}
\centering
\includegraphics[width=6cm,angle=-90]{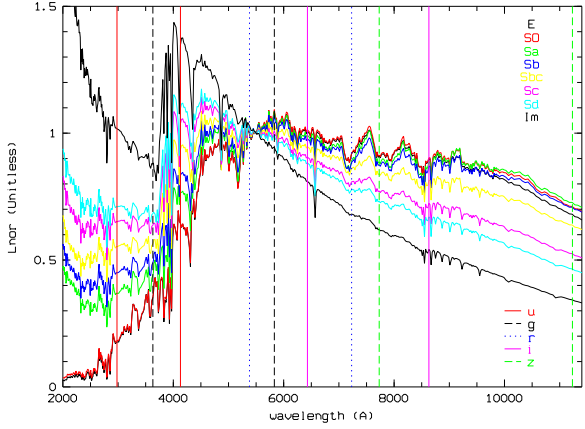}
\caption{
  Synthetic spectra for the eight typical galaxy types from P\'EGASE.2. The
  vertical lines denote the limits of the five SDSS filters (transmission
  below 1e-4 of the peak). (Emission lines are not included). The legend at
  the right defines colour used to plot each type of galaxy (top) and SDSS
  filter (bottom).}
\label{f1}
\end{figure}

Typical synthetic spectra corresponding to each of the eight Hubble types are
shown in Fig. \ref{f1}, with the location of the SDSS filters superimposed.
Each of these ``typical spectra'' corresponds to specific combination of
values of the astrophysical parameters (see section~\ref{most_signif}). The
SEDs produced by P\'EGASE have been normalized to the flux of a 50$\AA$
wavelength interval centered on 5500$\AA$. The elliptical and S0 galaxies have
very small differences, apparent at the two extremes of the wavelength range.
This implies small differences in colours but not necessarily in magnitudes
(which depend on their masses).

From Fig.~\ref{f1} it is obvious that the u filter is very important for the
comparison with real data since it is the one containing the discontinuity
around 4000$\AA$. However, the SDSS spectra do not cover the u band, so
photometry in this band cannot be synthesized. We refrain from using the SDSS
photometry for the u band because of the red leak in this
filter\footnote{http://www.sdss.org/dr4/products/images/index.html$\#$redleak},
which would render comparisons with synthetic data unreliable. This leak
produces erroneous magnitudes, especially for E and S0 types on
account of their large numbers of red stars.

In addition we avoid using the z filter in our comparison since 
its photometry also cannot be synthesized from the SDSS spectra, which terminate 
at shorter wavelengths than the z passband.

We therefore decided to base our comparison between the SDSS and P\'EGASE.2
data using the g, r, i filters only and, more specifically, the g--r and r--i
colours. However, the wavelength range of the SDSS spectra does not quite
extend to the bluest side of the g filter. For this reason, we cut the blue
end of this and created a new g filter starting at 3830$\AA$ instead of the
3630$\AA$ (table \ref{t1}). However, this change is in practice very small
since the transmission of the g filter is only 3\% of the peak transmission at
3830$\AA$ and drops very rapidly below that (e.g.\ it is only 0.5\% at just
10\AA\ lower). Furthermore, simulated photometry from the synthetic spectra
showed virtually no difference for the original and ``trimmed'' g band. The
published transmission curves of the SDSS filters depend on airmass and
whether a point or extended source is being observed.
We use those for extended sources and zero airmass. The
photometry is calibrated on the AB system, as used by SDSS \citep{fukugita}.  

\begin{figure}
\centering
\includegraphics[width=6cm,angle=-90]{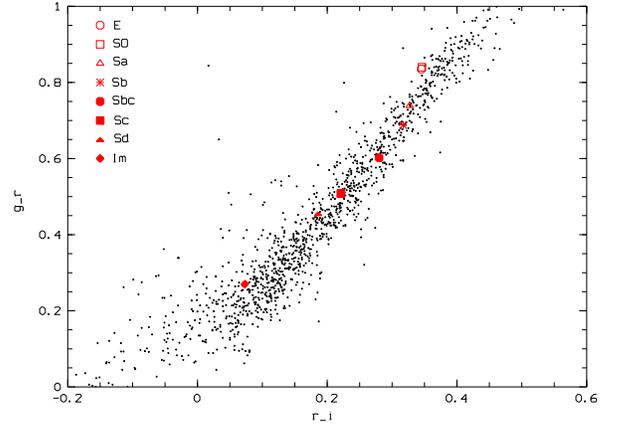}
\caption{Colour--colour (g$-$r vs.\ r$-$i) diagram of synthesized photometry of SDSS galaxy spectra (black) and synthetic photometry of the eight typical galaxy types generated from the 
  P\'EGASE models (red points).}
\label{f2}
\end{figure}

We synthesize photometry using the one-dimensional spectra from DR4, which are
supplied with additional analysis information, such as redshift and emission
line parameters. In order to select data suitable for our purposes, we
applied the following criteria: the galaxies should not be near a CCD edge nor
saturated, and they should not be very low SNR (the photometric error in all
bands should be less than 0.1 mag). Only spectra with redshifts below 0.01 are
retained, since the synthetic spectra of P\'EGASE.2 were produced at zero
redshift. These criteria resulted in a sample of 1292 galaxies. Their
synthesized photometry plus that for the eight typical galaxy types from
P\'EGASE.2 is shown in Fig.~\ref{f2}. This figure clearly shows that the
colours of the Im, Sd, Sc, Sbc, Sb and Sa types are generally in good
agreement with the colours of the observed spectra, although in the case of
S0 and E types the synthetic spectra seem to be slightly redder in g$-$r than
the SDSS spectra.

\section{The library of synthetic spectra}\label{library}

\subsection{The most significant parameters}\label{most_signif}

Each spectrum in our library is uniquely defined by a set of 17 astrophysical
parameters, plus the morphological type (E, S0, Sa, Sb, Sbc, Sc, Sd or Im).
The four most significant APs are: p1 and p2 of the star formation scenario
(($Mgas^{p_1}$)/p2); the infall timescale; the age of the galactic winds. The
age of the galactic winds is non-zero only for E and S0 galaxies. Note that
the Hubble type is not an independent parameter, as only certain ranges of the
APs are available for each type (as will be detailed later).

\begin{figure}
\centering
\includegraphics[width=6cm,angle=-90]{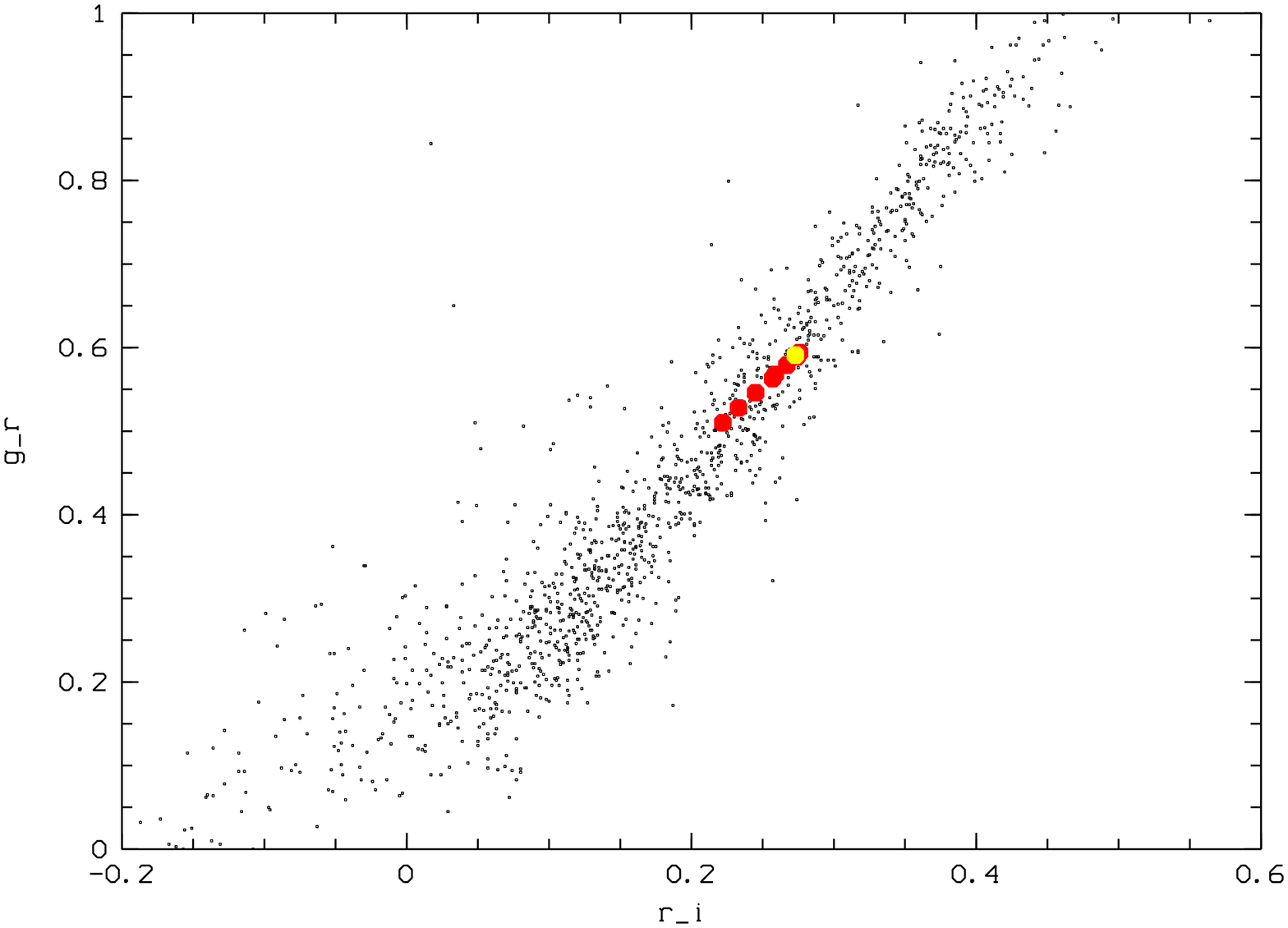}
\caption{Colour--colour (g--r vs r--i) diagram of synthesized photometry of SDSS galaxy spectra (black) and of synthetic P\'EGASE 
  spectra of the typical Sbc model (yellow) and the models of Sbc with
  different values of p1 (red). The largest g--r corresponds to p1=1 and the
  smallest g--r to p1=2.}
\label{f3}
\end{figure}

\begin{figure}
\centering
\includegraphics[width=6cm,angle=-90]{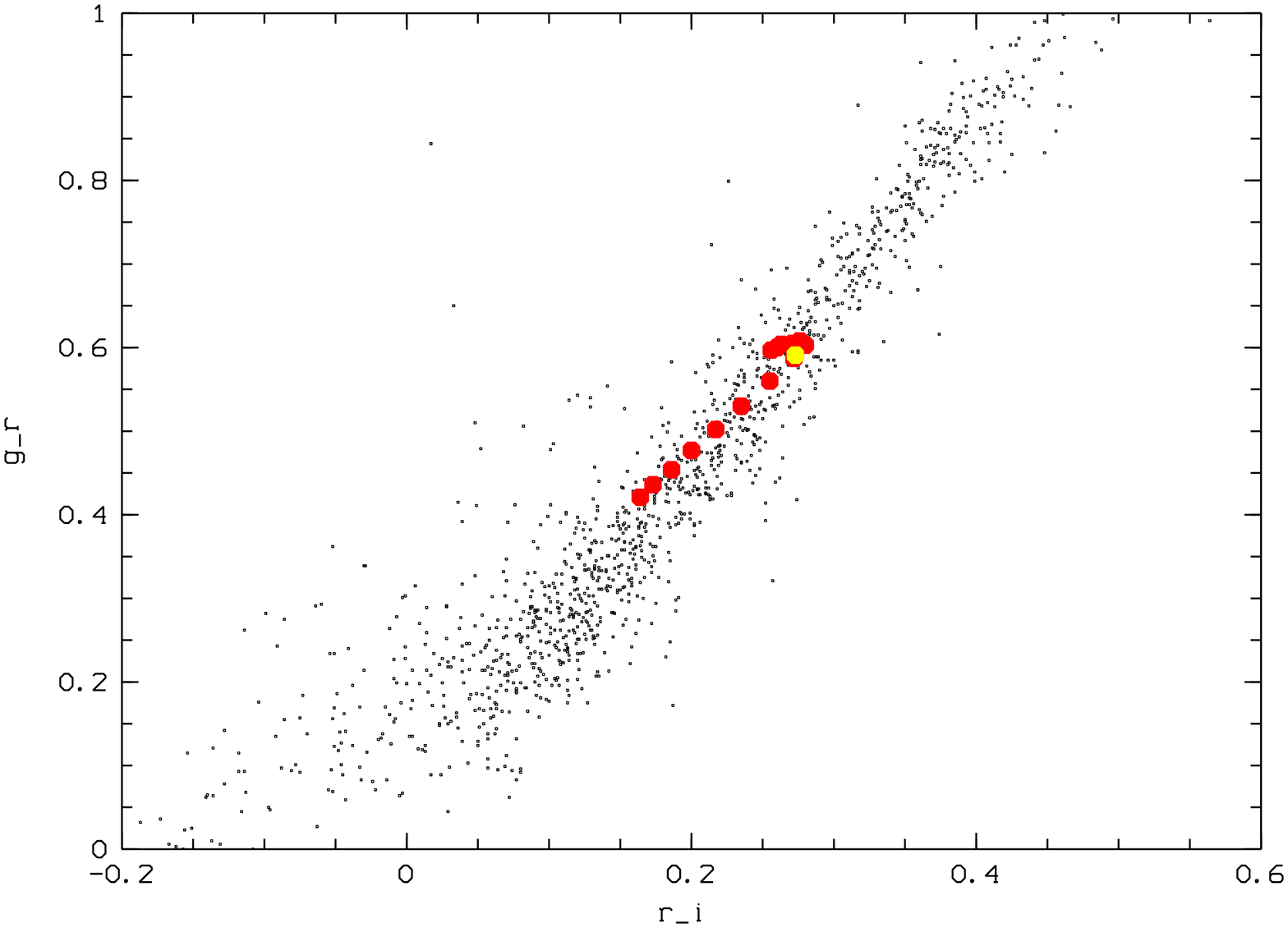}
\caption{Colour-colour (g--r vs r--i) diagram of synthesized photometry of SDSS galaxy spectra (black) and of synthetic P\'EGASE 
  spectra of the typical Sbc model (yellow) and the models of Sbc with
  different values of p2 (red). The largest g--r corresponds to p2=2000 Myr/$M_{\odot}$ and the
  smallest g--r to p2=20000 Myr/$M_{\odot}$.}
\label{f4}
\end{figure}

\begin{figure}
\centering
\includegraphics[width=6cm,angle=-90]{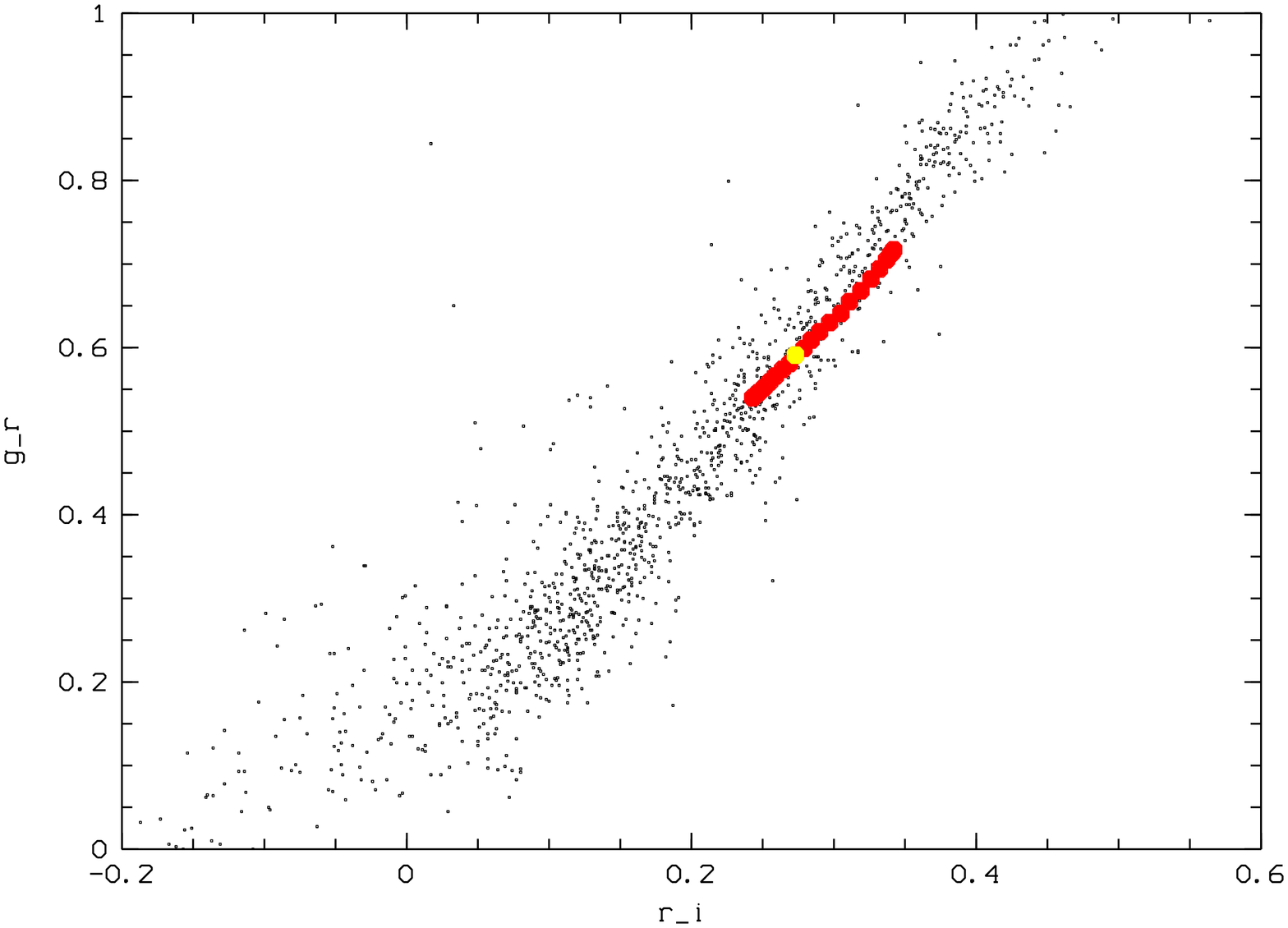}
\caption{Colour--colour (g--r vs r--i) diagram of synthesized photometry of SDSS galaxy spectra (black) and of synthetic P\'EGASE 
spectra of the typical Sbc model (yellow) and the models of Sbc with different values of infall timescale (red). The largest 
g--r corresponds to infall timescale=100My and the smallest g--r to infall timescale=10 Gyr.}
\label{f5}
\end{figure}

In order to investigate the influence of each of the parameters p1, p2 and
infall timescale to the integrated galaxy spectrum (SED), we modified the
parameters of the Sbc model (an intermediate type) over a range of values. In
the typical model for the Sbc type the values were 1, 5714 Myr/$M_{\odot}$ and 6000 Myr for
p1, p2 and infall timescale, respectively. In the modified models we vary p1
between 0.4 and 2, p2 from 100 to 20000 Myr/$M_{\odot}$ and infall from 100 to 10000 Myr. The
results are shown in Figs.~\ref{f3}--\ref{f5}.

\begin{figure}
\centering
\includegraphics[width=6cm,angle=-90]{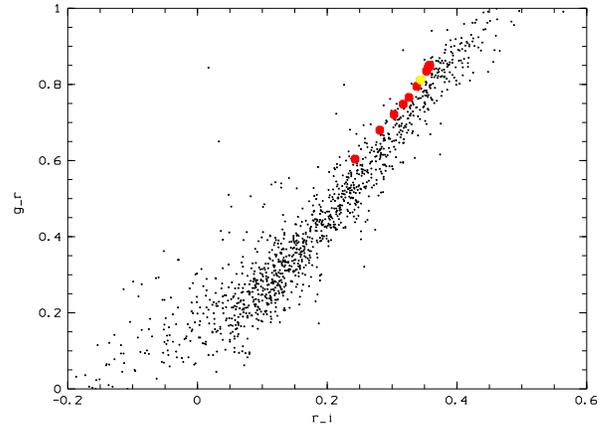}
\caption{Colour--colour (g--r vs r--i) diagram of synthesized photometry of SDSS galaxy spectra (black) and of synthetic P\'EGASE 
spectra of the typical E model (yellow) and the models of E with different values of age of galactic winds (red). The 
largest g--r corresponds to age of galactic winds=7.5Gy and the smallest g--r to 0.1 Gyr.}
\label{f6}
\end{figure}

To investigate the effect of the age of the galactic winds parameter we followed
the same procedure but now with the elliptical model. In the typical model for
the E type the age is 1~Gyr and we vary it between 0.1 and 7.5 Gyr
(Fig.~\ref{f6}).

From the figures we see that these four parameters have a major effect on the
colours. We performed similar analyses for other APs and concluded that they
had a much smaller impact on the data (in particular once the spectra are
reduced to the Gaia resolution). Therefore, the spectra in the present library
show variance only in these four APs.

By co-varying these four parameters and using all their combinations in each of
the eight typical models we are able to cover most of the variance we see in
the SDSS data in the colour--colour diagram. Generally, there is no clear
distinction between the colours of neighbouring Hubble types. In order to
have a knowledge of types in our library we decided (as a first working approximation) to only
retain those models which lie within a circle (in the colour-colour diagram)
centered on one of the eight typical types and with a radius equal to half of
the distance to the nearest neighbouring typical model. This is reasonable
since the models lie mostly on a one-dimensional surface (line) in the
colour--colour diagram. In this way upper and lower limits of the values of
the parameters were established for each type, although in this case
an overlap in APs (if not in colours) remains, as can be seen in table \ref{t2}.
This leaves a set of 888 synthetic spectra of known types of
galaxies (see section~\ref{regular_grid}).

The galaxy type can be considered as a 5th AP, although it is of a different
nature than the others, since it is needed to fully specify the spectrum and
constrain the range of values of the other four APs. In addition, when one
redshifts the spectrum to non-zero values of z, this quantity also becomes a
parameter (albeit not intrinsic to the source).

\subsection{Library of galaxy spectra over a regular grid of parameters}\label{regular_grid}

Applying the above procedures, we produced a library of 888 synthetic spectra
covering seven separate Hubble types (because we consider E and S0 as a single
type). The values of the four parameters of each type are given in table
\ref{t2}, while the values of the other input parameters of P\'EGASE.2 (kept
constant in all models) are given in table \ref{t3}. The models are plotted
in Fig.~\ref{f7}, where the simulated colours of the 888 synthetic spectra and
the 1292 SDSS spectra are compared. This first set of 888 synthetic spectra
was then calculated at five values of redshift: 0, 0.05, 0.1, 0.15, 0.2,
resulting in a total of 4440 spectra.

\begin{table*}
 \centering
 \caption {The four astrophysical parameter (AP) ranges for each Hubble type in
the regular library of P\'EGASE synthetic spectra. Note that the AP ranges for
each Hubble type partially overlap. The morphological type can be considered as
an additional (but non-independent) parameter, required to fully explain the
variance in the library. The final column (N) gives the number of spectra for
each type (which sum to 888). See the regular library grid in \citet{le2} for
comparison.}                                      
 \begin{tabular}{c c c c c c}          
 \hline\hline                        
  
Type & p1      & p2          & infall     & galactic winds & N  \\
     &         &(Myr/Msol)   &(Myr)       &(Gyr)           &    \\
 \hline
E-S0 & 0.6-1.5 & 100-1500    & 100-2500   & 0.1-7.5        & 327\\
Sa   & 0.8-1.5 & 500-2500    & 2500-3500  & none           & 10 \\
Sb   & 0.6-1.5 & 1500-6000   & 2000-4500  & none           & 25 \\
Sbc  & 0.4-1.5 & 2000-10000  & 4000-7000  & none           & 148\\
Sc   & 0.6-1.5 & 6000-14000  & 7000-10000 & none           & 68 \\
Sd   & 0.4-1.5 & 10000-18000 & 7000-10000 & none           & 65 \\
Im   & 1.0-2.0 & 14000-20000 & 7000-10000 & none           & 245\\
\hline
\end{tabular}
\label{t2}
\end{table*}
\begin{table*}
 \centering
 \caption {The values of the parameters of the P\'EGASE models which are kept
constant in the library \citep{fioc2}.}                                      
 \begin{tabular}{c c}          
 \hline\hline                        
Parameters & Values \\
\hline  
SNII Ejecta of massive stars                     & model B of \citet{woosley}\\
Stellar winds                                    & yes\\
Initial mass function                            & \citet{rana}\\
Lower mass                                       & 0.09 solar masses \\
Upper mass                                       & 120.00 solar masses\\
Fraction of close binary systems                 & 0.05\\ 
Initial metallicity                              & 0.00 \\
Metallicity of the infalling gas                 & 0.00\\
Consistent evolution of the stellar metallicity  & yes \\
Mass fraction of substellar objects              & 0.00\\
Nebular emission                                 & yes \\
Extinction                                       & disk geometry: inclination-averaged \\
                                                 & for Sa, Sb, Sbc, Sc, Sd and Im  \\
                                                 & spheroidal geometry for E-S0 \\
Age                                              & 13 Gyr  for E-S0,Sa, Sb, Sbc, Sc \& Sd \\
                                                 & 9 Gyr for Im \\
\hline
\end{tabular}
\label{t3}
\end{table*}

\subsection{Extension of the library to random values of parameters}

After producing the regular synthetic spectral grid (table \ref{t2}), we
proceed to produce synthetic spectra of galaxies with parameters selected
from a random distribution, in order to achieve a more continuous coverage in
colour space. Such grids permit more robust tests of parameter estimation
algorithms than do regular grids. Each parameter is selected independently
from a uniform distribution over the parameter ranges in the regular grid. We
used this approach to generate 5500 models. In doing this we keep
approximately the ratios between the Hubble types as in the regular grid.
Because the parameter ranges for each galaxy type in Table \ref{t2} show some overlap,
a random draw may produce a set of parameters which fits into more than one
Hubble type category. To remove this ``degeneracy'' we again apply the circle
removal method we used in section~\ref{most_signif}. This results in a
``non-degenerate'' sample of 2709 spectra.

\begin{figure}
\centering                    
\includegraphics[width=6cm,angle=-90]{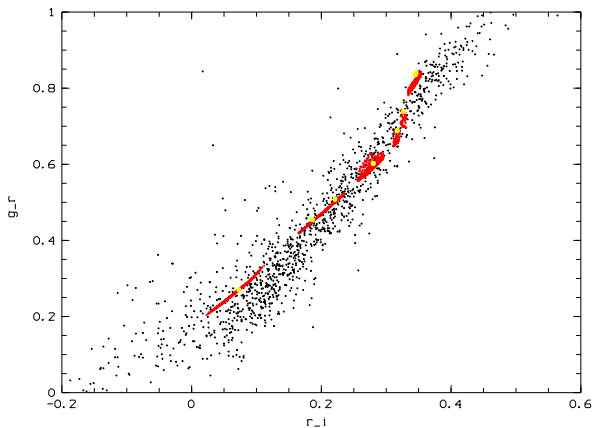}
\caption{ Colour--colour (g--r vs.\ r--i) diagram of synthesized photometry of
SDSS galaxy spectra (black) and of synthetic P\'EGASE spectra of the 8 typical
models of P\'EGASE.2 (yellow). Moving from the lower left to the upper right
part of the diagram we encounter types from Im to E. The red dots along both
sides of the typical models represent the spectra of both the regular and
random library.}
\label{f7}
\end{figure}

A comparison of the simulated colours of the synthetic spectra (888 regular
grid plus 2709 random grid, at zero redshift) with the colours of SDSS spectra
is shown in Fig. \ref{f7}. One sees that the new set of spectra is in very
good agreement with the SDSS data, except for the small differences in the E
and S0 galaxies.

In summary, we have produced a library of 7149 synthetic galaxy spectra (888
spectra of the regular grid for 5 values of redshift and 2709 of the random
grid at zero redshift) which can be used as an initial library of unresolved
galaxy spectra for assessing the possibilities of galaxy classification and
parametrization with Gaia. This library was created at the resolution of the
BaSeL 2.2 stellar library 
(gradually changing from 8\,\AA\ at 2500\,\AA\ to 50\,\AA\ at 10\,500\,\AA),
which is not quite high enough for the Gaia simulation software (which
requires 10\,\AA). Therefore, we linearly interpolated our spectra in order
to resample the spectra to 10\,\AA\ over the wavelength range of
2500--10\,500\,\AA. Higher resolution spectra will be produced in future work
using the High-spectral Resolution code P\'EGASE-HR \citep{le1}.

\section{Simulated Gaia spectra}

\begin{figure}
\centering                    
\includegraphics[width=6cm,angle=-90]{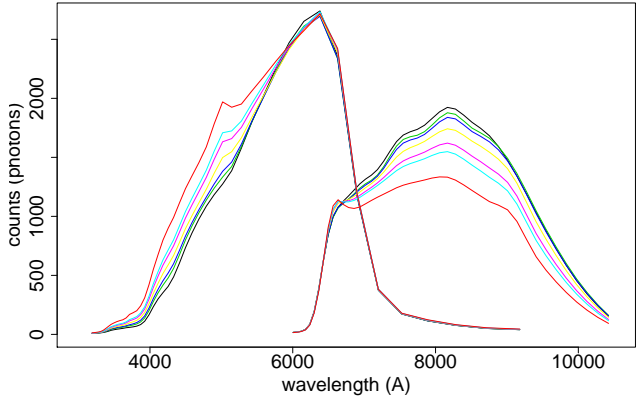}
\caption{ The simulated BP and RP spectra of the synthetic spectra for the
eight typical galaxy types from P\'EGASE.2. Black, green, blue, yellow,
magenta, light blue and red denote galaxies of type E, Sa, Sb, Sbc, Sc, Sd and
Im respectively.}
\label{f8}
\end{figure}

The Gaia spectrophotometer is a slitless prism spectrograph comprising blue
and red channels (called BP and RP respectively) which operate over the
wavelength ranges 3300--6800\,\AA\ and 6400-10\,500\,\AA\ respectively. BP
and RP spectra were simulated for all 7149 library spectra using the simulator
developed by \citet{brown}. Each of BP and RP is simulated with 48 pixels,
whereby the dispersion varies from 30--290\,\AA/pix and 60--150\,\AA/pix
respectively. We artificially reddened each spectrum with a standard
interstellar extinction law with R=3.1, for regular values of $A_{V}$ from 0
to 10 for the regular library, and for 10 random values of $A_{V}$ uniformly
distributed in $log(1+A_{V})$ for the the random library. Noise was added to
all spectra, which includes the source Poisson noise, background Poisson noise
and CCD readout noise. This is done for five different source G-band
magnitudes (15, 17, 18, 19 and 20). For the following classification tests we
use only the sample at G=18. In Fig. \ref{f8} we present the simulated BP and RP spectra
for the eight typical synthetic spectra of galaxies. 

\section{Classification \& Parametrization}

In the present work we use classification Support Vector Machines (SVMs)
(C-classification) to determine morphological types and regression SVMs
($\epsilon$-regression) to estimate the various astrophysical parameters. We
use the libsvm library of \citet{libsvm} implemented in the \verb+e1071+
package in the R statistics package.\footnote{{\tt http://www.r-project.org}} A
brief description of the SVMs is given in the Appendix of this paper. An
accessible introduction to SVMs can be found in \citet{bennett00}. For a more
technical introduction, the tutorial by \citet{burges98} is recommended.

\subsection{Galaxies at zero redshift}

\subsubsection{Classification of the morphological type}\label{classify}

We now try to classify the set of Gaia-simulated galaxy spectra, at G=18 with
zero redshift, into the seven Hubble types. This subset of the library
includes characteristic noise and a wide range of interstellar extinction
(from 0--10 mag in $A_{v}$). It comprises 9691 spectra. This we divide at
random into two subsets: 4846 for training the SVM classifiers and 4845 for
evaluating their performance. As is recommendable with many machine learning
methods, we first normalized the data by scaling each input (pixel) to have
zero mean and unit standard deviation.

\begin{figure}
\centering                    
\includegraphics[width=6cm,angle=-90]{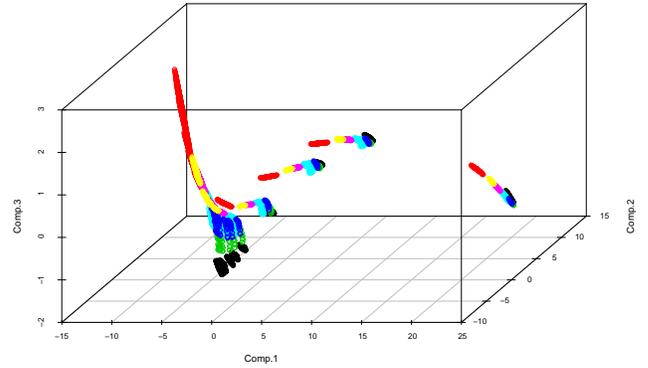}
\caption{ The 9691 simulated Gaia galaxy spectra with z=0 plotted as their
projections onto the first three Principal Components. Black, green, blue,
light blue, magenta, yellow and red denote galaxies of type E, Sa, Sb, Sbc, Sc,
Sd and Im respectively.}
\label{f9}
\end{figure}

For the purpose of visualizing the data set only, we perform a Principal
Components Analysis (PCA) on the set of 9691 96-dimensional Gaia spectra.
The first three Principal Components describe 78.25\%, 20.44\% and 1.02\% of
the data variance respectively (i.e.\ 99.71\% together).\footnote{Note that,
because each input dimension has already been normalized to have zero mean
and unit variance, a considerable fraction of the total variance is already
accounted for.} In Fig. \ref{f9} we plot the data in projection onto the
first three PCs. This diagram, plus the fact that the first three PCs explain
almost all of the variance in the data, suggest that a good classification
should be possible (the data have an intrinsic low dimensionality).

\begin{table}
      
 \centering 
 \caption {Galaxy classification with the SVM. The confusion matrix for the
training set for galaxies at z=0. Columns indicate the true class, row the
predicted ones.}                                   
 \begin{tabular}{l | c c c c c c c}          
 \hline\hline                        
  
Type & E-S0 & Sa & Sb & Sbc & Sc & Sd & Im \\
 \hline
E-S0 & 1799  & 0    & 0  & 0   & 0   & 0   & 0   \\
Sa   & 0     & 1366 & 0  & 0   & 0   & 0   & 0   \\
Sb   & 0     & 0    & 53 & 5   & 0   & 0   & 0   \\ 
Sbc  & 0     & 0    & 0  & 134 & 0   & 0   & 0   \\
Sc   & 0     & 0    & 0  & 0   & 830 & 0   & 0   \\ 
Sd   & 0     & 0    & 0  & 0   & 0   & 347 & 1   \\
Im   & 0     & 0    & 0  & 0   & 0   & 0   & 311 \\

\hline
\end{tabular}
\label{t4}
 \centering
 \caption {As Table~\ref{t4} but for the test set.}
 \begin{tabular}{l | c c c c c c c}          
 \hline\hline                        
  
Type & E-S0 & Sa & Sb & Sbc & Sc & Sd & Im \\
 \hline
E-S0 & 1798 & 0    & 0  & 0   & 0   & 0   & 0   \\
Sa   & 0    & 1329 & 0  & 0   & 0   & 0   & 0   \\
Sb   & 0    & 0    & 44 & 0   & 0   & 0   & 0   \\ 
Sbc  & 0    & 0    & 4  & 137 & 0   & 0   & 0   \\
Sc   & 0    & 0    & 0  & 1   & 797 & 0   & 0   \\ 
Sd   & 0    & 0    & 0  & 0   & 0   & 394 & 6   \\
Im   & 0    & 0    & 0  & 0   & 0   & 3   & 324 \\

\hline
\end{tabular}
\label{t5}
\end{table}
                        
The results of training and testing the SVM classifier on the full 96-pixel
spectra are shown in Tables \ref{t4} and \ref{t5}. We see that there are very
few misclassifications: only 6 and 14 in the training and testing set
corresponding to an error of 0.12\% and 0.29\% respectively. While these
results are very promising, it must be recalled that the way the library has been
constructed avoids class overlap in the SDSS g$-$r, r$-$i colour space,
which surely eases separation in the 96-dimensional BP/RP colour space.

\subsubsection{Regression of astrophysical parameters}

\begin{figure*}[t]
  \setlength{\unitlength}{1cm}
\begin{picture}(18,15)
\put(0,15){\includegraphics{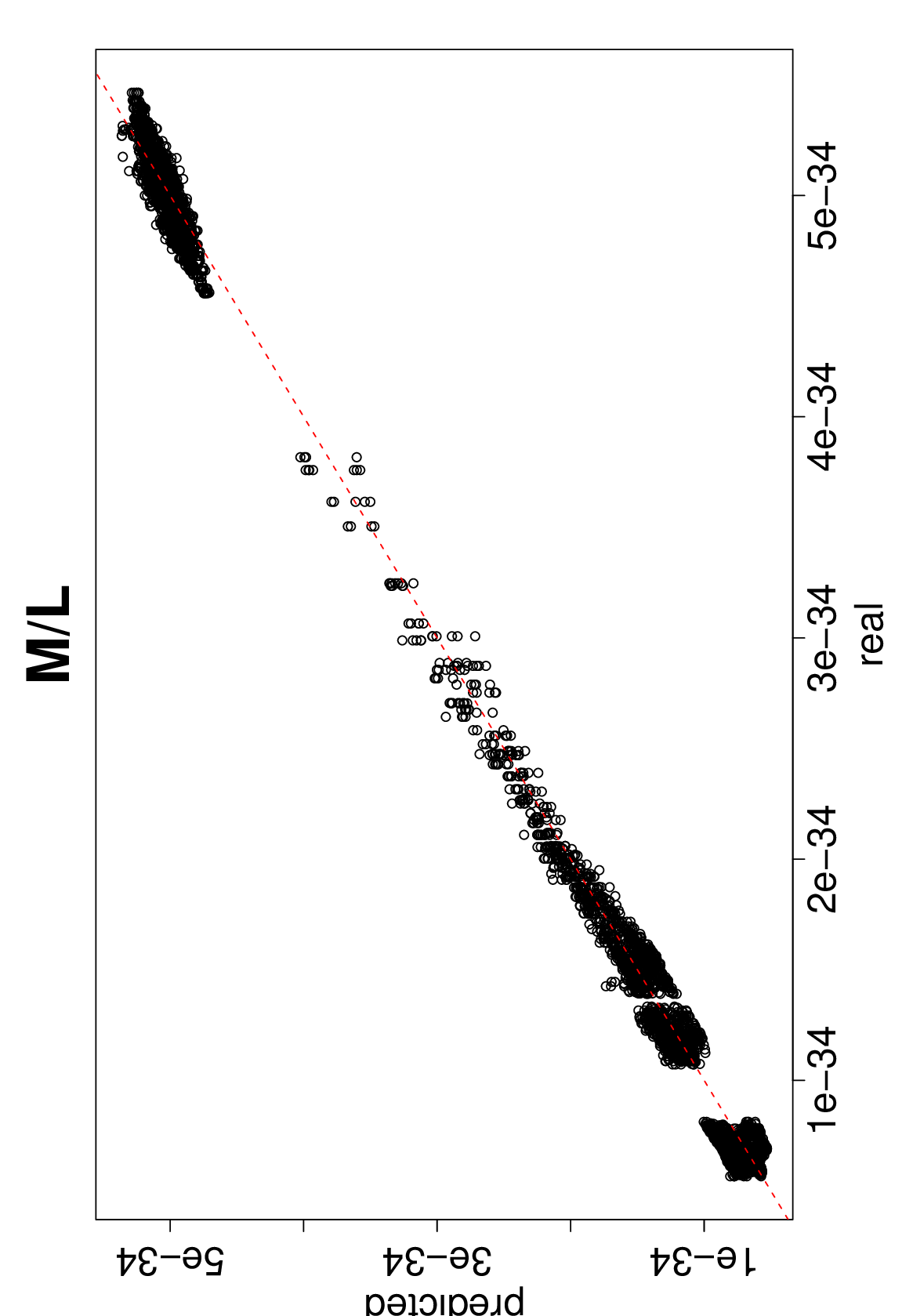}}
\put(0,10){\includegraphics{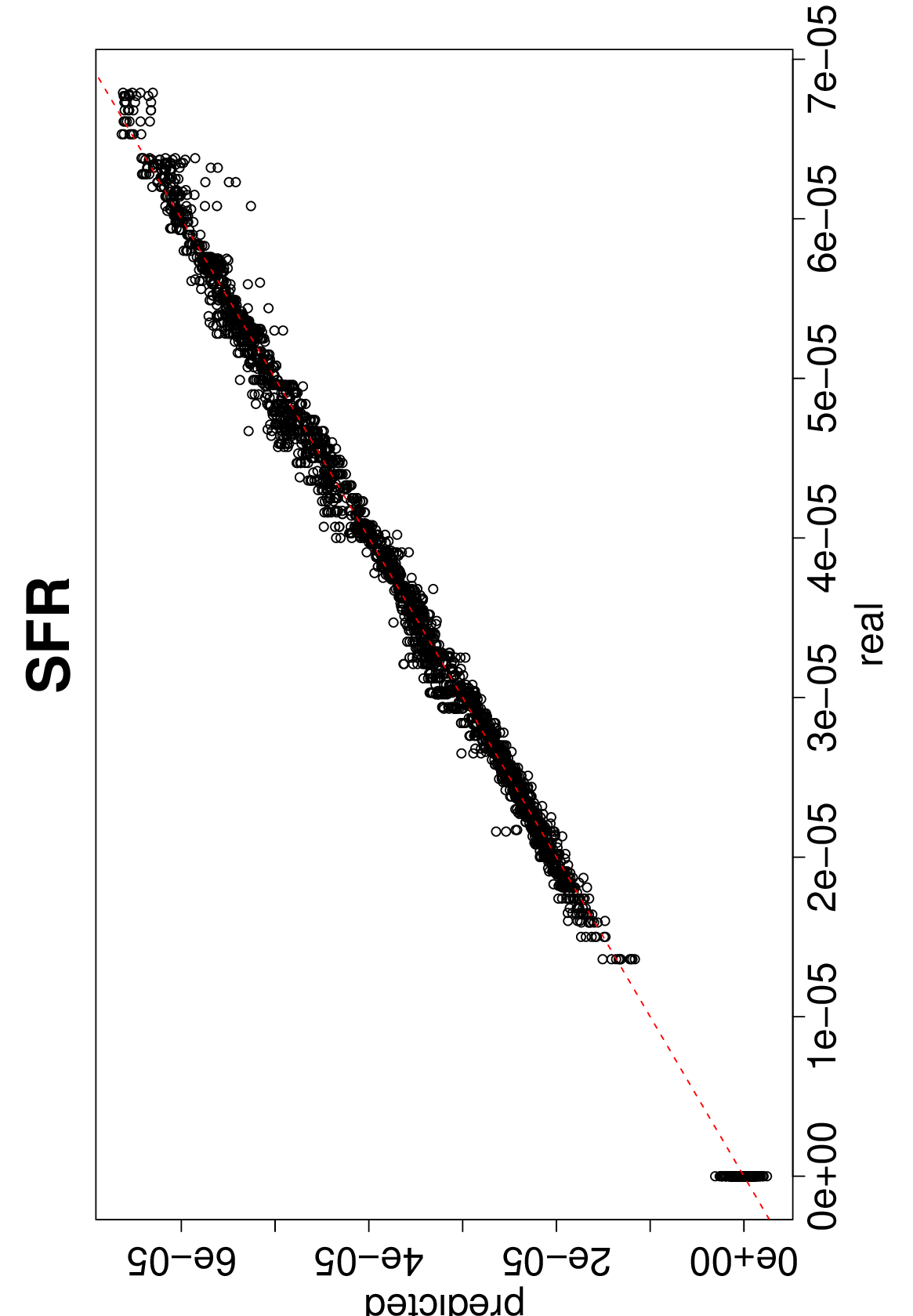}}
\put(0,5){\includegraphics{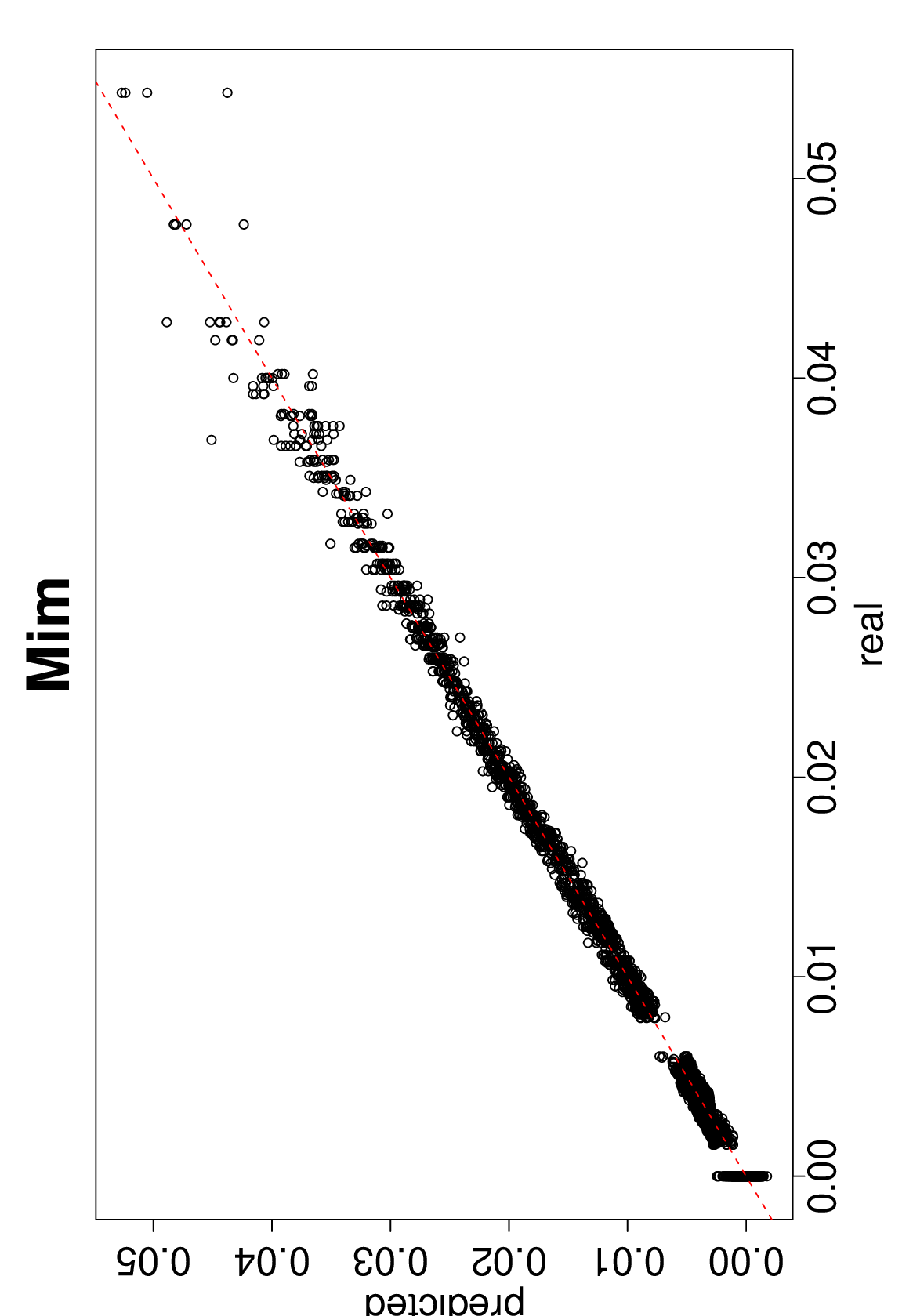}}
\put(6,15){\includegraphics{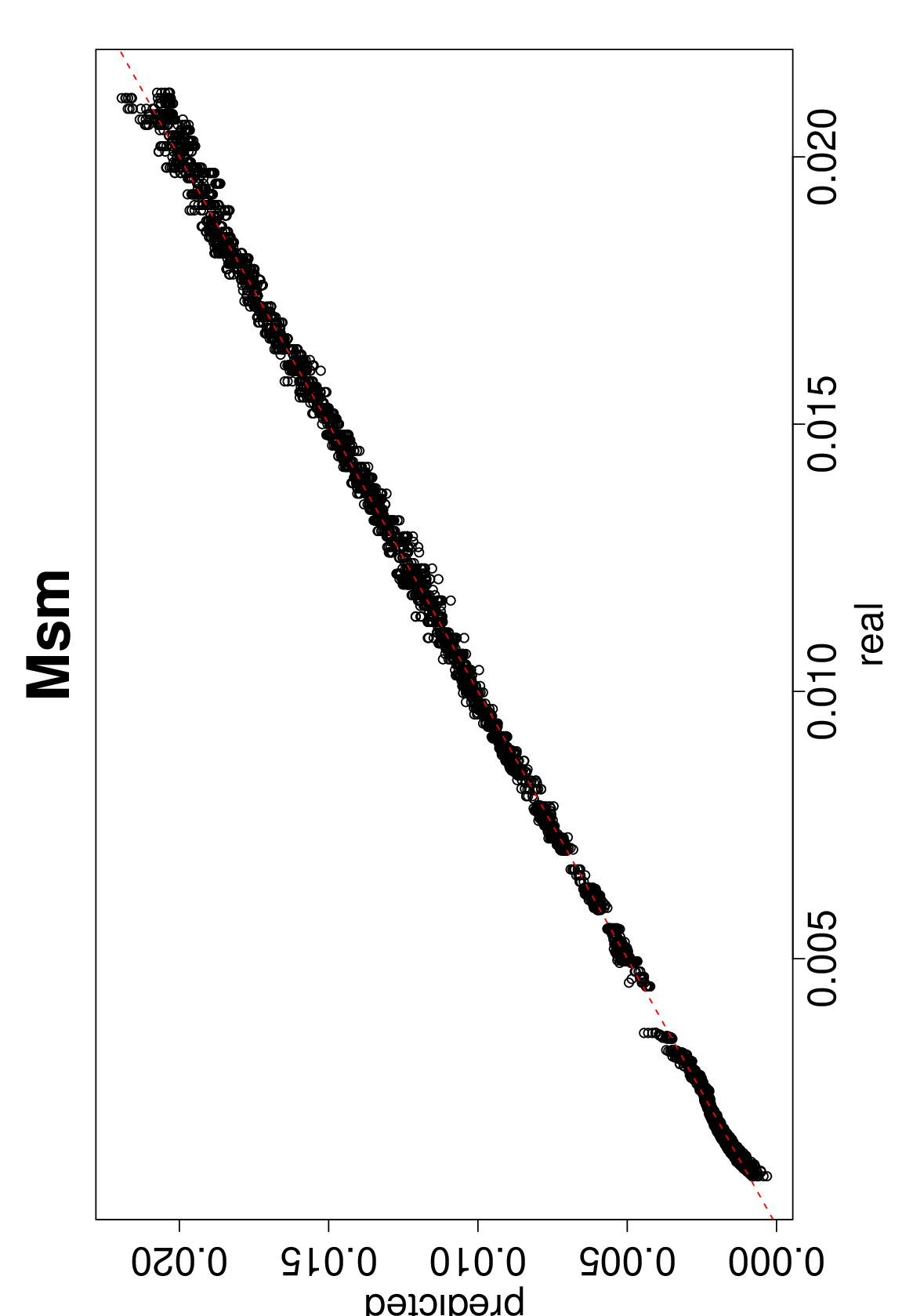}}
\put(6,10){\includegraphics{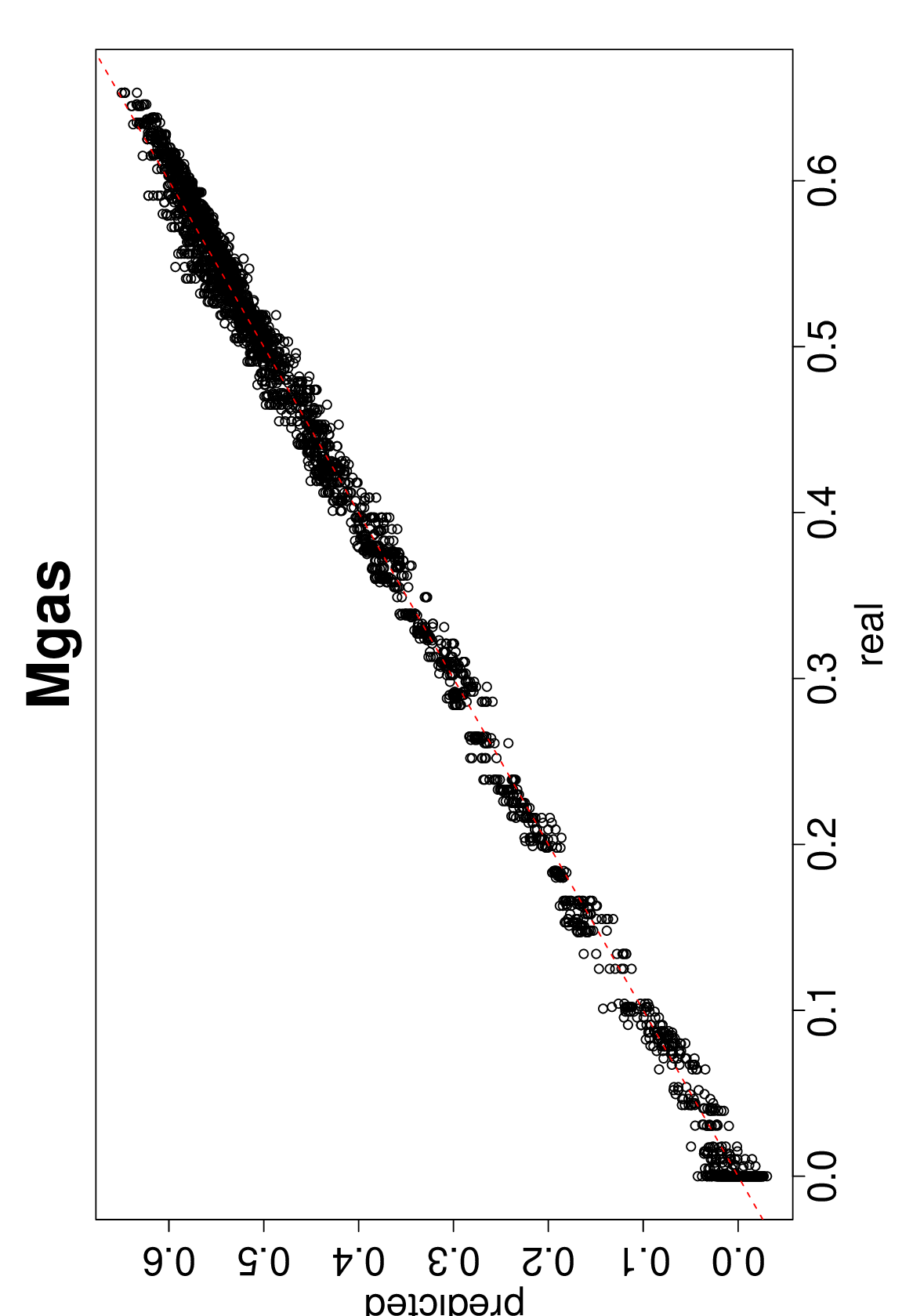}}
\put(6,5){\includegraphics{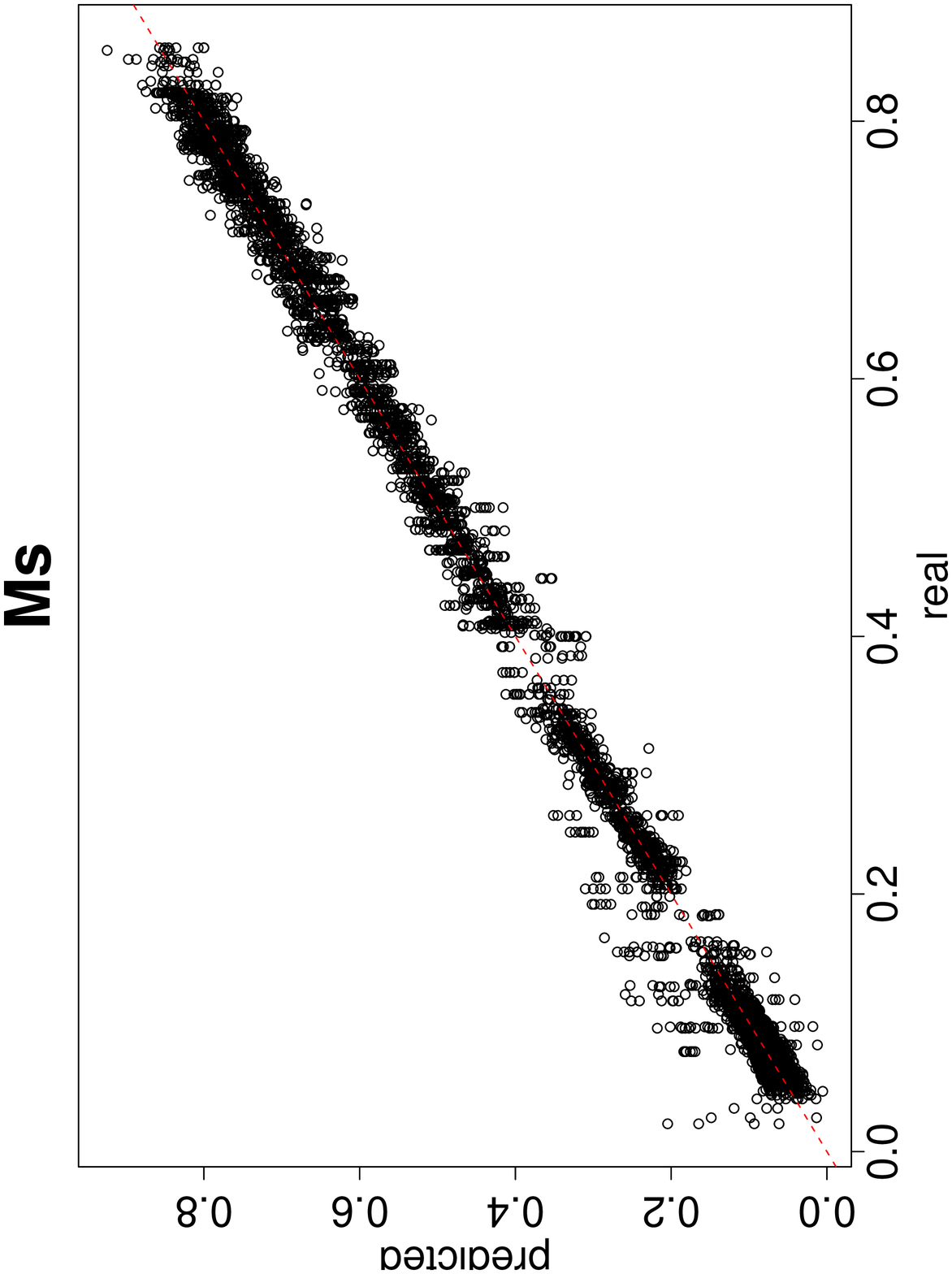}}
\put(12,15){\includegraphics{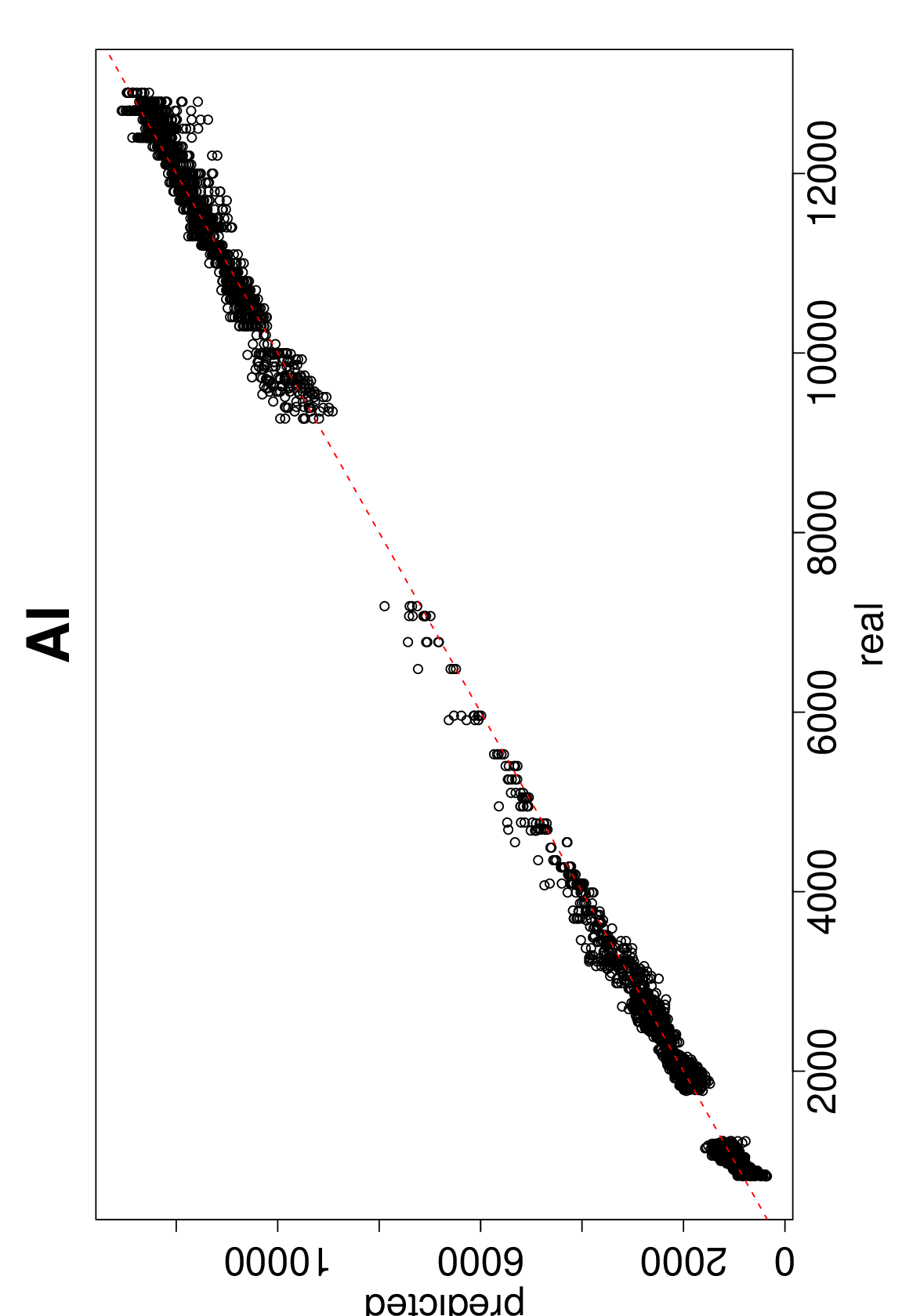}}
\put(12,10){\includegraphics{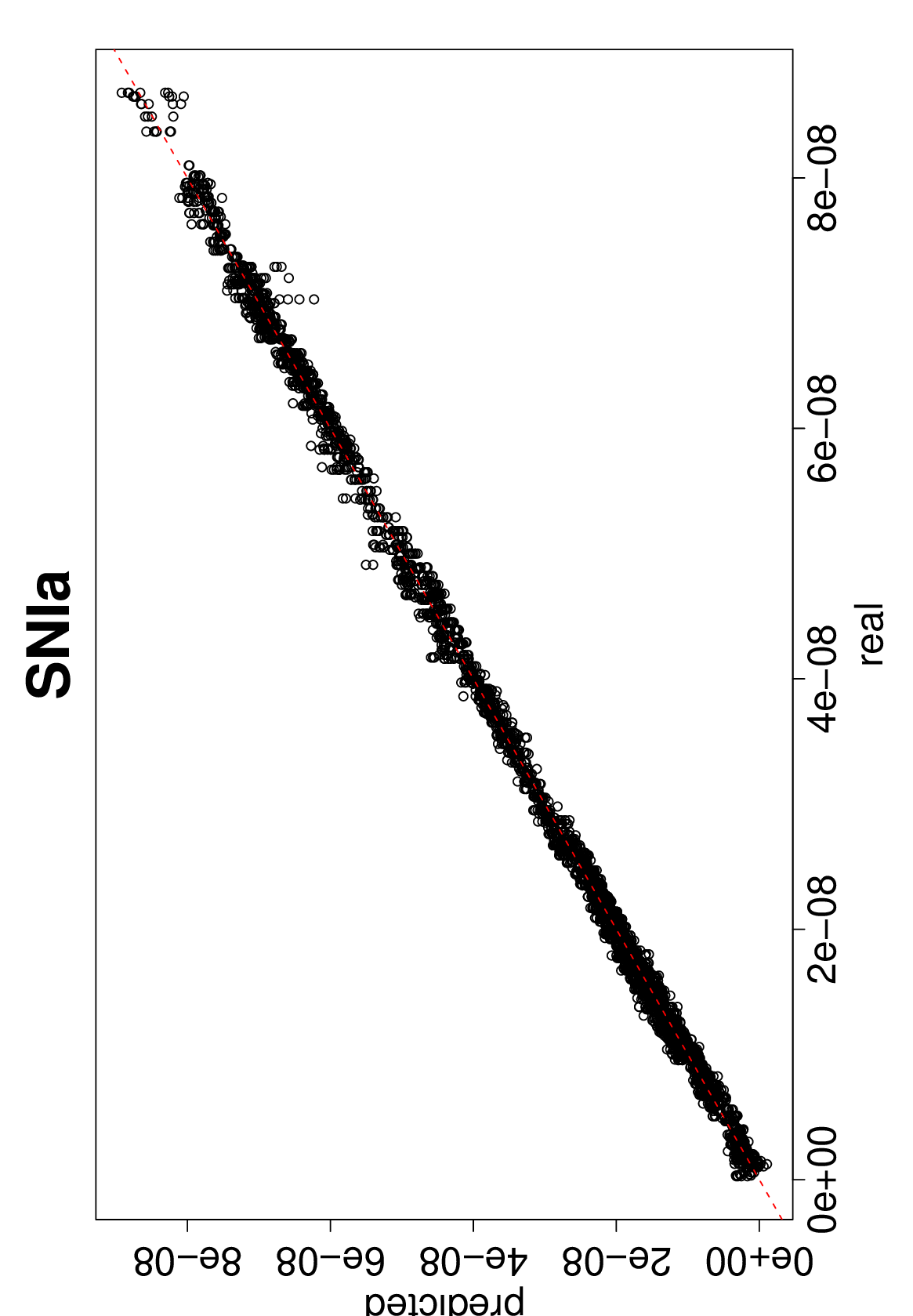}}
\put(12,5){\includegraphics{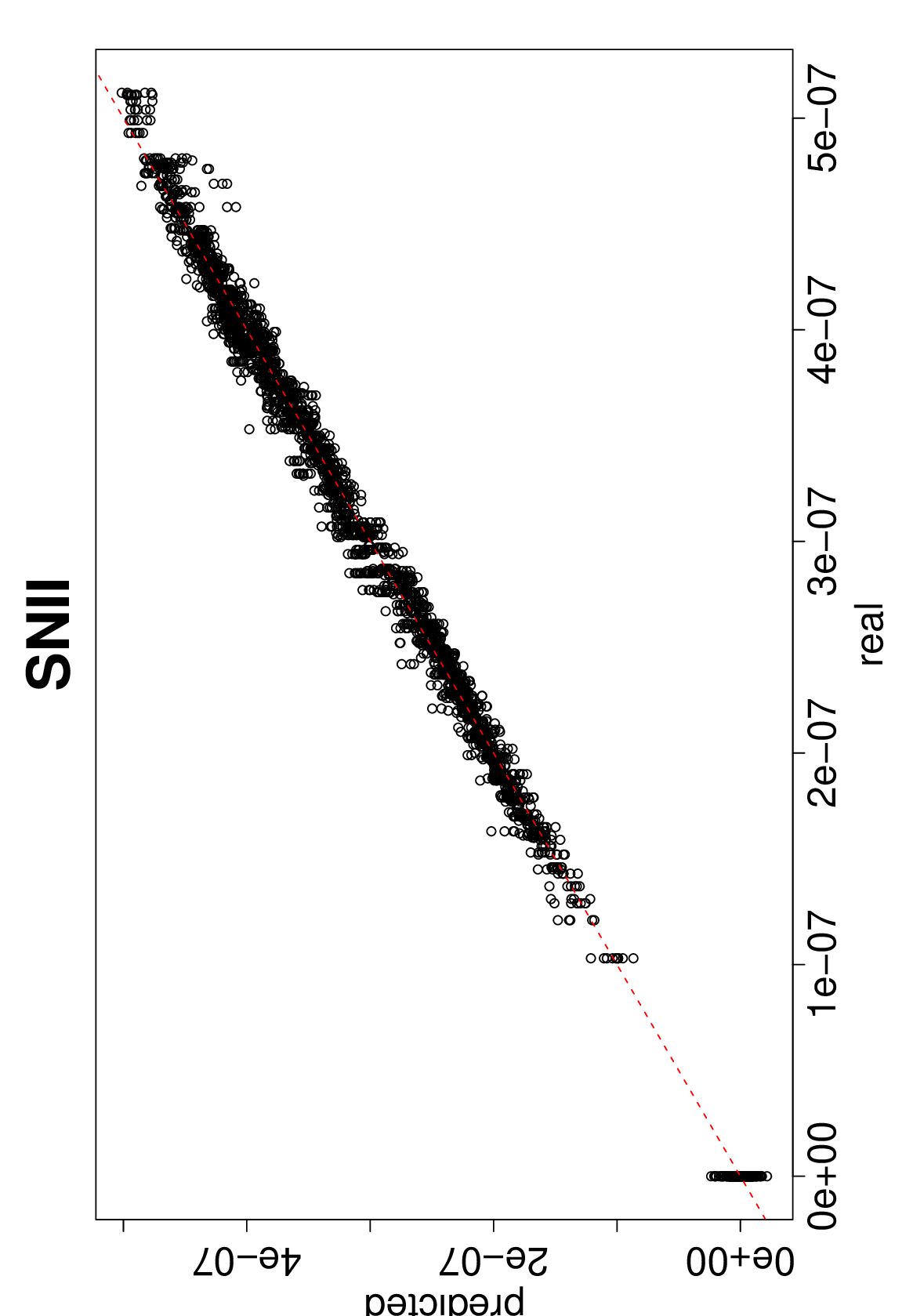}}
\end{picture}
\caption{
  Galaxy parameter estimation performance. For each of the nine APs we plot
  the predicted vs.\ true AP values for the test set. The red line indicates
  the line of perfect estimation. The summary errors are given in
  Table~\ref{t6}.}
\label{f10}
\end{figure*}

In addition to simulating an output spectrum, P\'EGASE.2 also derives 18
output astrophysical parameters for each galaxy. Of course, by construction we know that our
synthetic spectra are uniquely defined by five parameters (p1, p2, infall
timescale, age of the galactic winds and the Hubble type), so there can only
be five equivalent independent parameters amongst these 18. Nonetheless, it
would be useful to predict them directly. Here we build SVM regression models
to separately predict the nine most significant ones (listed in Table
\ref{t6}). For each model we train on a randomly selected set of 4846 spectra
and evaluate performance on the remaining 4845. In Fig.~\ref{f10} we present
the true and the SVM-predicted values of each parameter on the test set. Table
\ref{t6} summarizes this by giving the mean of the difference between the true
and predicted values for each parameter (which measures the systematic error)
as well as the RMS residual (which measures the total scatter). The plots and
table indicate that we can predict the parameters to good accuracy and
precision, i.e.\ the systematics are very small and the RMS error is a small
fraction of the typical values.

\begin{table*}
 \centering
 \caption {Summary of the performance of the SVM regression models for
predicting the nine APs listed. The sample is for zero redshift but for
interstellar extinction ($A_{v}$) varying from 0 to 10\,mag. The second and
third columns list the mean and RMS errors respectively. The final column gives
the number of support vectors in the SVM model.}
 \begin{tabular}{l c c c}          
 \hline\hline                        
  
Astrophysical Parameter                                  & mean(real-predicted)/mean(real) & sd(real-predicted)/mean(real) & SVs   \\
 \hline
mass to light ratio (M/L)                                & -1.03e-2             & 3.78e-2            & 97    \\
normalized star formation rate (SFR)                     & -3.35e-3             & 3.97e-2            & 2285  \\
metallicity of interstellar medium (Mim)                 & -2.85e-3             & 8.77e-2            & 345   \\ 
metallicity of stars averaged on mass (Msm)              & -3.64e-4             & 2.17e-2            & 3544  \\
normalized mass of gas (Mgas)                            &  4.52e-3             & 4.29e-2            & 190   \\ 
normalized mass in stars (Ms)                            &  3.22e-4             & 5.48e-2            & 1639  \\
mean age of stars averaged on bolometric luminosity (Al) &  1.45e-3             & 3.22e-2            & 3566  \\
normalized SNIa rate (SNIa)                              &  9.69e-4             & 3.43e-2            & 376   \\
normalized SNII rate (SNII)                              & -6.04e-4             & 3.81e-2            & 2247  \\
\hline
\end{tabular}
\label{t6}
\end{table*}

\subsection{Galaxies with redshift}

\subsubsection{Regression of redshift and classification of morphological type}

We now enlarge the subset of the library we used in the previous tests by
adding the same galaxies at four nonzero values of redshift, specifically
0.05, 0.1, 0.15, 0.2. The library for z=0 includes 9691 galaxies as described above.
For each nonzero redshift there are 9757
giving a total sample of 48\,719 galaxies. (Recall that this includes each
galaxy simulated at 11 regular values of $A_{v}$.) We now build another
morphological type classification model as done in section~\ref{classify}, now
with 6719 galaxies in the training set and 42\,000 galaxies for testing set.

\begin{figure}
  \centering \includegraphics[width=6cm,angle=-90]{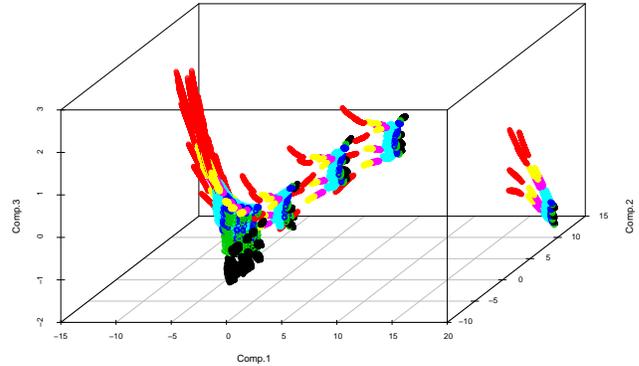}
 \caption{ The 48\,719 simulated Gaia galaxy spectra with nonzero redshift
plotted as their projections onto the first three Principal Components. Black,
green, blue, light blue, magenta, yellow and red denote galaxies of type E, Sa,
Sb, Sbc, Sc, Sd and Im respectively.}
\label{f11}
\end{figure}

We again applied a PCA to the data. This time the first three Principal
Components describe 76.01\%, 21.63\% and 1.02\% of the data variance
respectively (i.e.\ 98.6\% together), very similar to before. The
corresponding PCA-project plot is Fig.~\ref{f11}. Comparing to Fig.~\ref{f9} we
can see how the redshift spreads out the previous loci of types.
The performance of the SVM classifier is summarized in Tables \ref{t7} and
\ref{t8}. The performance is good considering the added complexity introduced
by the redshift variations (and the corresponding increase in the sample
size). The misclassification errors are 0.13\% and 0.98\% corresponding to 9
and 411 galaxies for the training and the testing data respectively.

\begin{table}
 \centering
 \caption {Galaxy classification with the SVM. The confusion matrix for the
training set for galaxies at z=0.0, 0.05, 0.1, 0.15, 0.2. Columns indicate the
true class, row the predicted ones.}
 \begin{tabular}{l | c c c c c c c}          
 \hline\hline                        
  
Type & E-S0 & Sa & Sb & Sbc & Sc & Sd & Im \\
 \hline
E-S0 & 2512 & 0    & 0  & 0   & 0    & 0   & 0   \\
Sa   & 0    & 1828 & 0  & 0   & 0    & 0   & 0   \\
Sb   & 0    & 0    & 74 & 2   & 0    & 0   & 0   \\ 
Sbc  & 0    & 0    & 1  & 183 & 1    & 0   & 0   \\
Sc   & 0    & 0    & 0  & 0   & 1115 & 0   & 0   \\ 
Sd   & 0    & 0    & 0  & 0   & 0    & 536 & 4   \\
Im   & 0    & 0    & 0  & 0   & 0    & 1   & 462 \\

\hline
\end{tabular}
\label{t7}
 \centering                                      
 \caption {As Table~\ref{t7} but for the test set.}
 \begin{tabular}{l | c c c c c c c}          
 \hline\hline                        
  
Type & E-S0 & Sa & Sb & Sbc & Sc & Sd & Im \\
 \hline
E-S0 & 15473 & 0     & 0   & 0    & 0    & 0    & 0    \\
Sa   & 0     & 11647 & 0   & 0    & 0    & 0    & 0    \\
Sb   & 17    & 0     & 344 & 113  & 0    & 0    & 0    \\ 
Sbc  & 0     & 0     & 83  & 1084 & 23   & 0    & 0    \\
Sc   & 0     & 0     & 8   & 39   & 6971 & 7    & 0    \\ 
Sd   & 0     & 0     & 0   & 0    & 1    & 3149 & 50   \\
Im   & 0     & 0     & 0   & 0    & 0    & 70   & 2921 \\

\hline
\end{tabular}
\label{t8} 
\end{table}

\begin{figure*}[t]
\setlength{\unitlength}{1cm}
\begin{picture}(18,10)
\put(0,10){\includegraphics{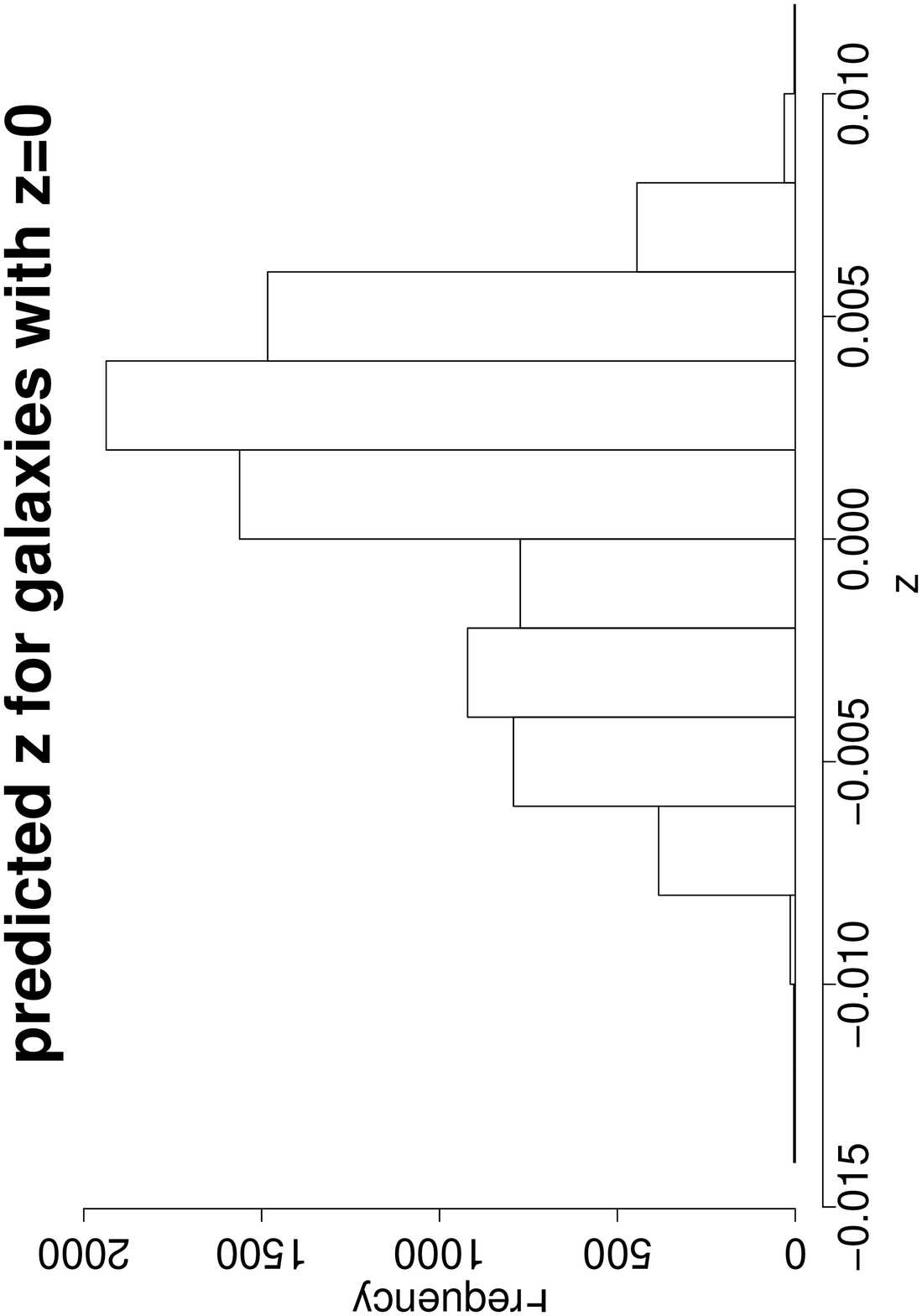}}
\put(6,10){\includegraphics{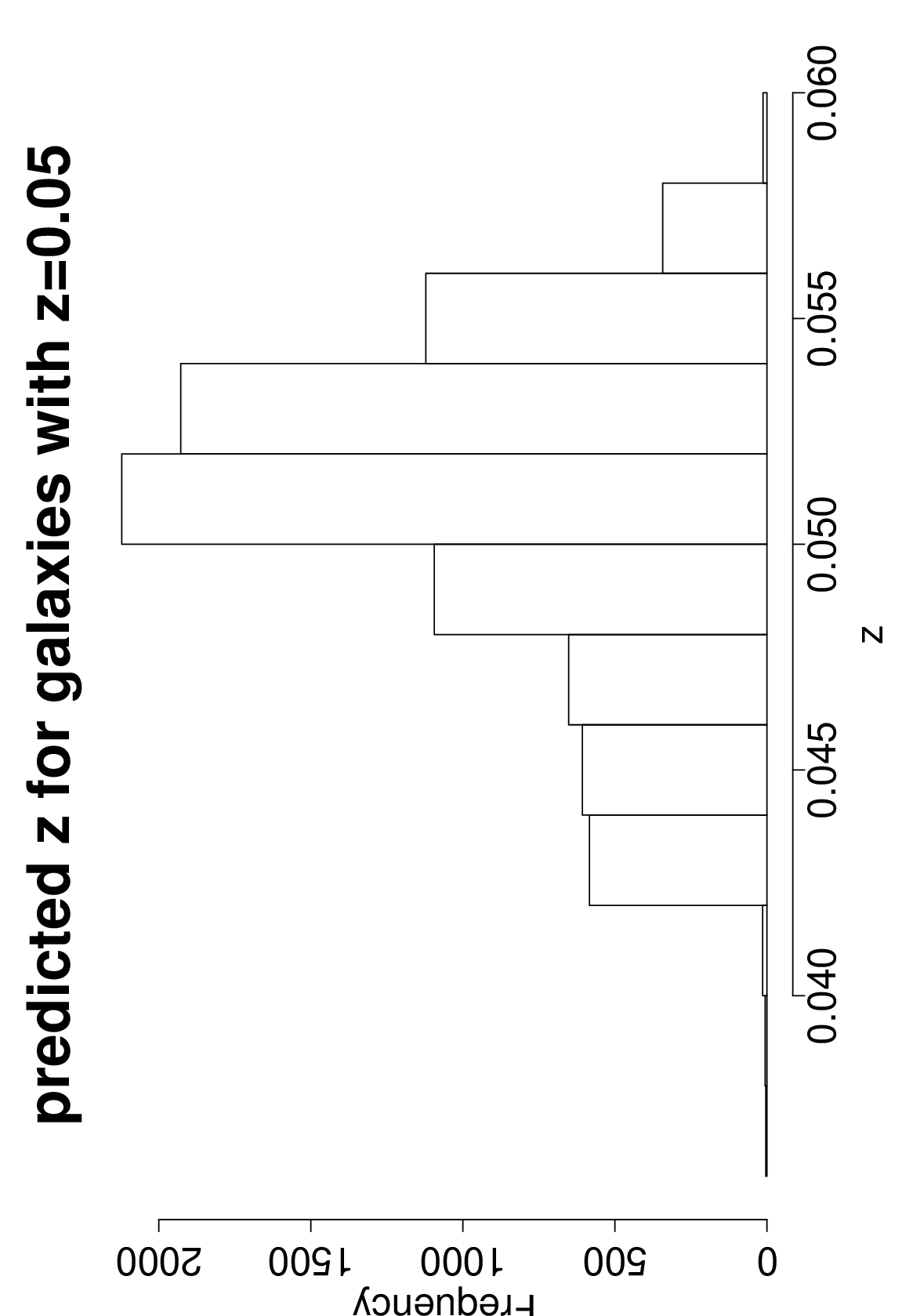}}
\put(12,10){\includegraphics{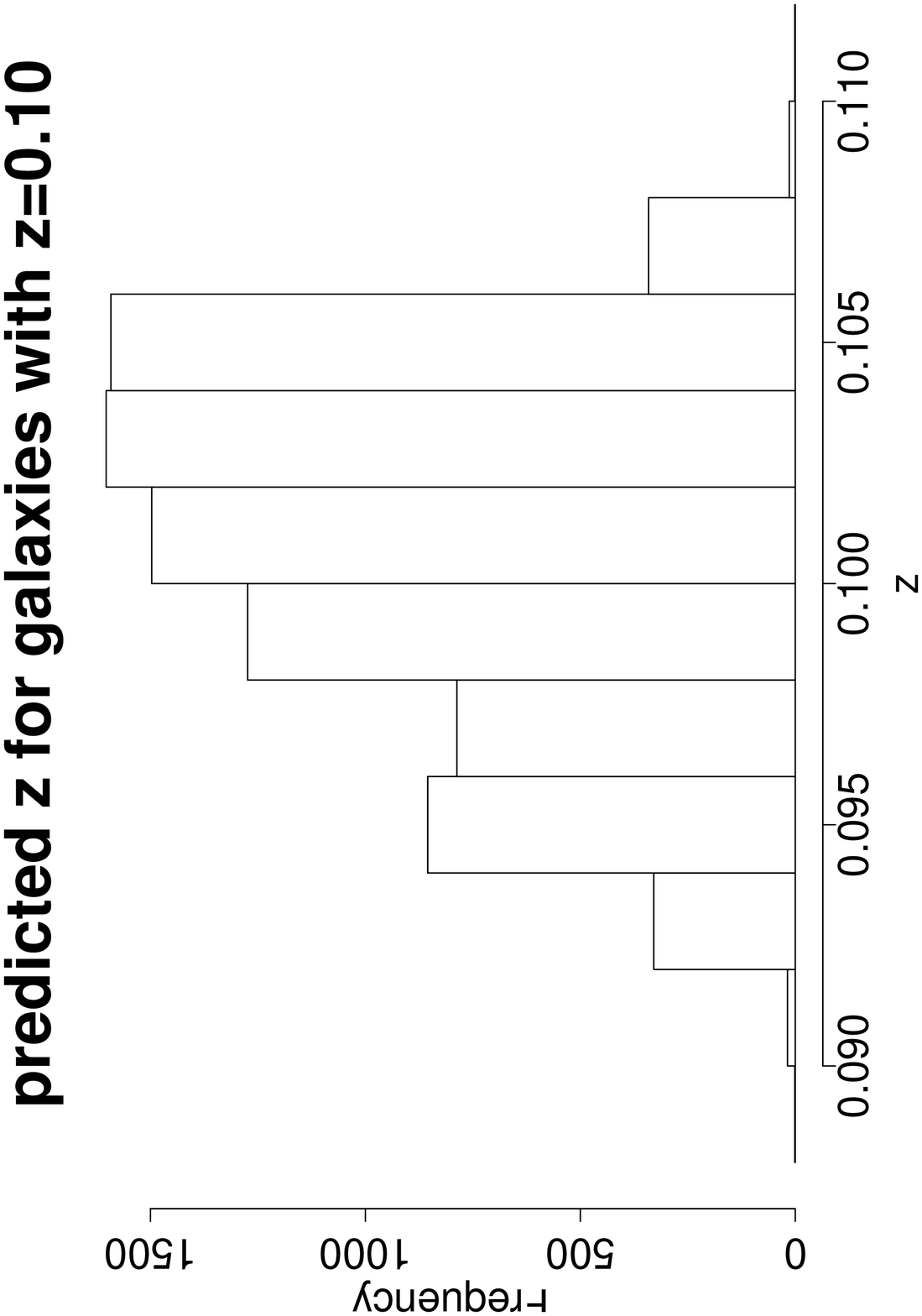}}
\put(4,5){\includegraphics{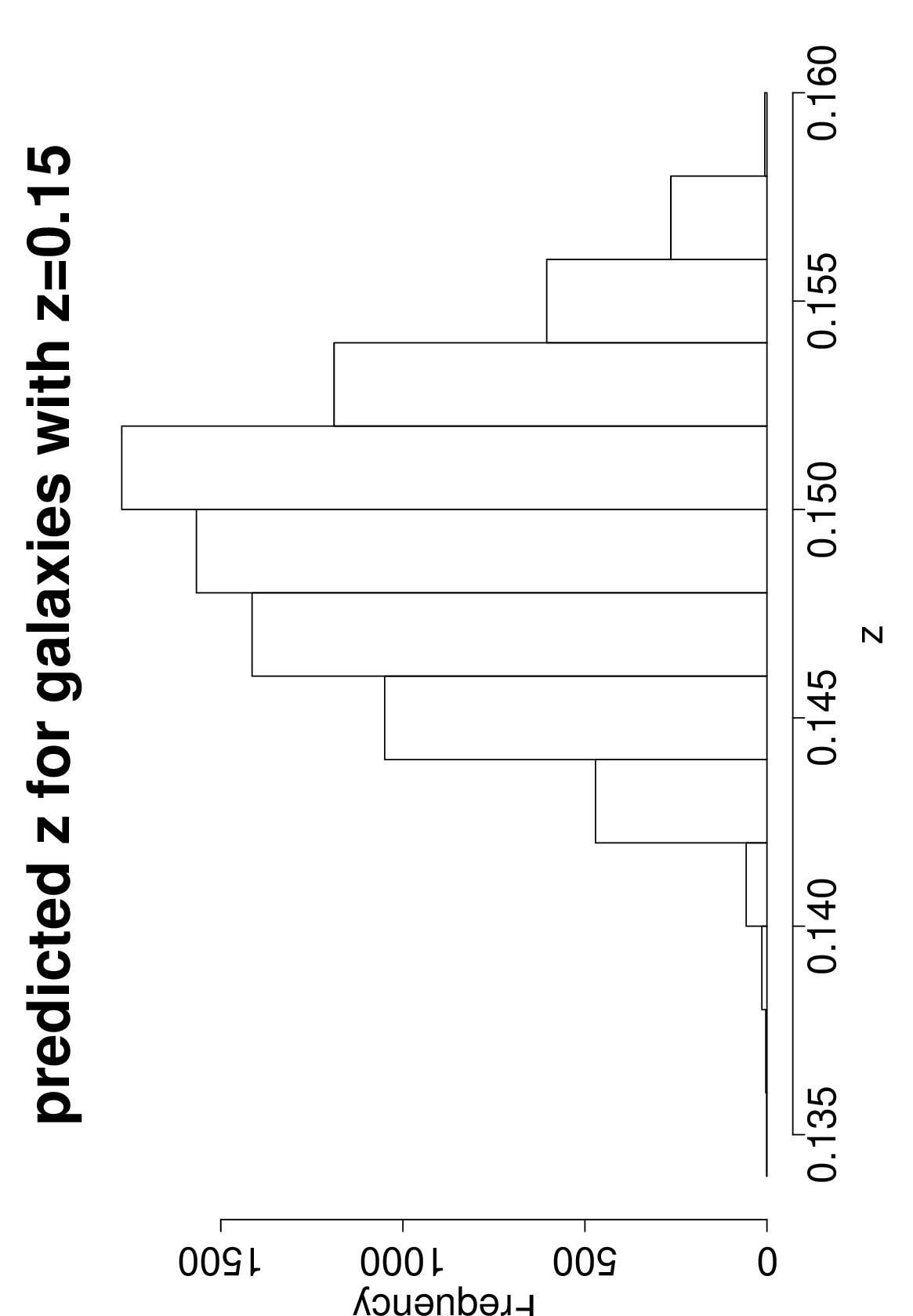}}
\put(10,5){\includegraphics{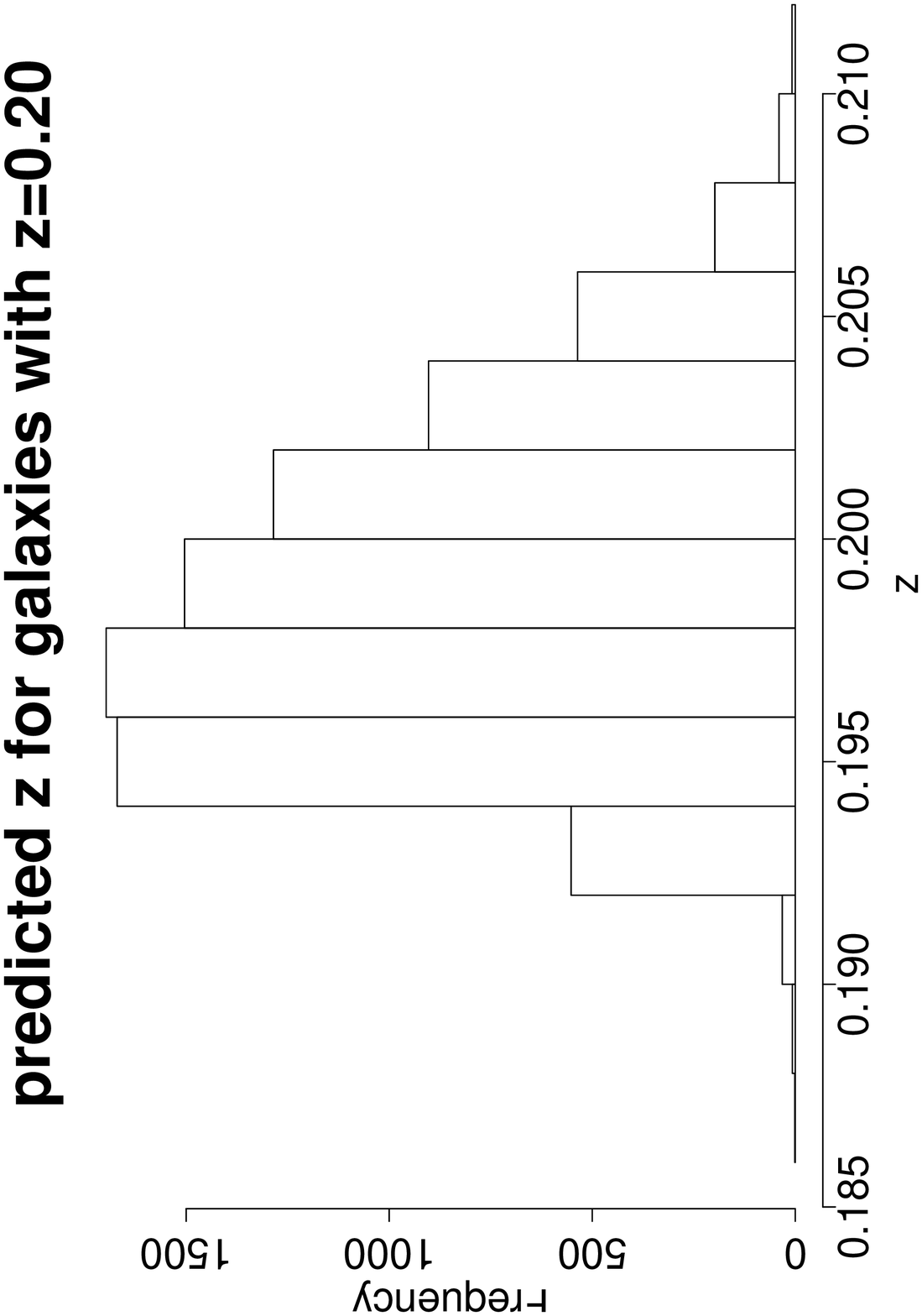}}
\end{picture}
\caption{Distribution of predicted values of redshift shows separately for the five true values of redshift (z=0, 0.05, 0.1, 0.15 and 0.2)}
\label{f12}
\end{figure*}

In practice we may want to first reduce spectra to the rest frame, for which
we require an estimate of the redshift. Therefore, we also set up a SVM
regression model to predict redshift, using the same training and test sets.
The predicted values of redshift for each of the five true redshift values are
presented in Fig. \ref{f12}. We do not expect very good performance here,
because the SVM is having to learn the effect of redshift based on just five
different values.

\section{Discussion and conclusion}

We have used the P\'EGASE.2 galaxy evolution model and the observational data
from SDSS to create an extended grid of synthetic galaxy spectra. Using these
we have identified the relevant astrophysical parameters and their relevant
ranges which provide a realistic galaxy spectra of known morphological type.
This was done specifically by comparing the colours of our library spectra
with those synthesized from SDSS spectra. We found small deviations between
the two colour loci for redder galaxies -- where the ellipticals are found --
which might be due to the fact that SDSS spectra are obtained in a small aperture
(fibre diameter) while P\'EGASE spectra are representative of the whole
galaxy. We also see that the observed sample has a considerably larger spread
in the colour--colour diagram than the library spectra, which probably has
observational reasons (photometric errors) as well as theoretical ones
(insufficient cosmic variance in the galaxy models). That is, it may
partially reflect the complicated nature of galaxy formation and evolution,
although the overall agreement between the two is good.

To achieve a better agreement between the observational and
synthesized libraries we will further investigate the influence of the
various P\'EGASE.2 parameters, especially those that were kept constant for
this release of the library. On the other hand, due to the narrow redshift
range ($z < 0.2$) explored here, evolution factors are minimized. At higher
redshifts, synthetic spectra will be computed by simultaneously applying
cosmological k-corrections and evolution e-corrections to z=0 templates.

Among the existing libraries of observed spectra, the most complete
and homogeneous is the SDSS, since it covers a significant part of the whole
sky and it goes fainter than the expected detection limit of Gaia. We
therefore aim to produce a suitable set of synthetic spectra covering as
much as possible of the SDSS colour range and we plan further comparisons in
our future work.

Adding phenomena such as the galaxy mergers is a challenging
hypothesis, but we believe that at the low redshifts Gaia will observe, this
is not such an important or frequent mechanism of galaxy evolution. On the
other hand, starburst galaxies are more frequent at small redshifts and we
intend to enrich our library with this type of galaxy.

First results of SVM for classification and parametrization of the library are
quite promising. In particular, the first indications are that Gaia will be
able to produce a wealth of information for a large statistical sample of
galaxies. After constructing a more complete library of spectra we will be
able to perform more tests and construct a classifier able to treat more
realistic and complete simulations of galaxy spectra.

\section{Acknowledgments}
The authors (the Greek team) would like to thank the Greek General Secretariat
of Research and Technology (GSRT) for financial support.

P. Tsalmantza would also like to thank the Max-Planck-Institut f\"ur
Astronomie (MPIA) and Institut d'Astrophysique de Paris (IAP) for their
support and hospitality.

Funding for the Sloan Digital Sky Survey (SDSS) has been provided by the Alfred
P. Sloan Foundation, the Participating Institutions, the National Aeronautics
and Space Administration, the National Science Foundation, the U.S. Department
of Energy, the Japanese Monbukagakusho, and the Max Planck Society. The SDSS
Web site is http://www.sdss.org/.

The SDSS is managed by the Astrophysical Research Consortium (ARC) for the
Participating Institutions. The Participating Institutions are The University
of Chicago, Fermilab, the Institute for Advanced Study, the Japan Participation
Group, The Johns Hopkins University, the Korean Scientist Group, Los Alamos
National Laboratory, the Max-Planck-Institute for Astronomy (MPIA), the
Max-Planck-Institute for Astrophysics (MPA), New Mexico State University,
University of Pittsburgh, University of Portsmouth, Princeton University, the
United States Naval Observatory, and the University of Washington.

\clearpage
\appendix
\section{Support Vector Machines}

Support Vector Machines (SVMs) \citep{vapnik95} are supervised machine
learning methods for data classification. In their basic form they achieve a
linear classification between two classes by defining an optimal hyperplane which
separates members of the two classes. If the classes are separable then there
generally exists an infinite number of hyperplanes which achieve this.
The SVM optimal plane is defined as that plane which maximises the margin
between the opposing class members nearest to the boundary. That is, unlike
many other classifiers which use all of the data to define the boundary, SVMs
take the (arguably more reasonable) approach of using just those points nearest
to the boundary. It has been demonstrated that this gives rise to a
more robust and more accurate classifier under general conditions.

In most non-trivial problems, however, the classes are not linearly separable.
In these cases, just those points which lie on the wrong side of the
hyperplane -- the so-called support vectors -- enter into the total
classification error. By minimizing this error -- which also measures the
distance of the support vectors from the plane -- we define the optimal
separating plane, i.e.\ with the fewest misclassifications (and preferentially
of those which lie closer to the plane).

In the general case, the classes are not even marginally linearly separable
(consider the XOR problem) so a linear classifier, no matter how optimal, is
useless. SVMs address this issue by using kernels to project the data into a
higher dimensional space. For example, with a polynomial kernel we take
square, cubic etc.\ combinations of the original data to form additional
dimensions and then apply the (linear) SVM classifier in this higher
dimensional space. With many other kernels, however, this projection is only
carried out implicitly. This approach can be thought of as nonlinearity by
preprocessing, with the kernel overcoming the well known ``curse of
dimensionality''. In the present work we use the radial basis kernel
\begin{equation}
K(x_i - x_j)=exp(-\gamma||x_i-x_j||^{2})
\label{SVM_kernel}
\end{equation}
where $x_i$ and $x_j$ are two input vectors (e.g.\ spectra). The
classification of a new vector $x_i$ is then given by a function
\begin{equation}
f(x_j) = \sum_{i}^{i=N} y_i \alpha_i K(x_i - x_j)
\label{SVM_model}
\end{equation}
where $y_i \in (-1,1)$ denotes the two classes, and a classification is made
by applying a threshold, e.g.\ $f(x_j) > 0.0 \Rightarrow$ class 1. The
${\alpha_i}$ are the parameters of the model which are determined by the model
training ($i$ counts over the $N$ support vectors). SVMs have a very important
property, namely that the error function is strictly convex, so it
has a unique global solution which can be found in polynomial time with
standard optimizers (it is a linearly constrained quadratic programming
problem).

This is in marked contrast to neural networks, for example, in which the optimizers
converge on a local minimum and we can only be guaranteed to find
the global minimum via an exhaustive search. Furthermore, with a sigmoidal
kernel SVMs are equivalent to neural networks but with the additional
advantage that the SVM automatically determines the neural network
architecture (number weights).

The SVM model incorporates regularization via the specification of a
hyperparameter, $C$, which defines the width of a margin around the separating
hyperplane. The wider this margin (larger $C$), the more data vectors 
which fall into it. These are all considered support vectors and so all enter the
error equation. Thus with a larger $C$ there is a higher penalty attached to
errors, i.e.\ less regularization.\footnote{$C$ is actually the upper bound on
$\alpha_i$, specifically $0 \leq \alpha_i \leq C$ and $\sum_i \alpha_i y_i =
0$ (two of the constraints in the error minimization). Thus a small $C$
implies smaller $\alpha_i$ in equation~\ref{SVM_model} which in turn implies
smoother functions equivalent to more regularization.}

The other hyperparameter in the model is $\gamma$ (equation~\ref{SVM_kernel}).
Both $\gamma$ and $C$ must be determined by the user. Prior information may
help, but in practice one carries out a rigorous search over a two-dimensional
grid to ``tune'' the SVM. We did this using 4-fold cross validation,
iterating over grids of increasing density.

SVMs can also be used for regression. Instead of a hyperplane and a margin
about it, regression SVMs fit a line with a tube of radius $\epsilon$
encompassing it. Data vectors which are less than a distance $\epsilon$ from
the line are considered to be correctly fit, that is, the support vectors are
only those points outside of the tube. Thus the $\epsilon$
hyperparameter controls the degree of regularization. The specific error
function we use is the mean squared error on the predictions, with
the regularization again being introduced via the constraints in the
optimization (with Lagrangian multipliers). All of the kernel and optimization
machinery applies equally to these models, so that nonlinear regression can
also be achieved.


\begin{thebibliography}{22}
\expandafter\ifx\csname natexlab\endcsname\relax\def\natexlab#1{#1}\fi

\bibitem[{{Armand} \& {Milliard}(1994)}]{FOCA2000}
{Armand}, C. \& {Milliard}, B. 1994, \aap, 282, 1

\bibitem[{Bennett \& Campbell(2000)}]{bennett00}
Bennett, K.~P. \& Campbell, C. 2000, SIGKDD Explor. Newsl., 2, 1

\bibitem[{{Brown}(2006)}]{brown}
{Brown}, A. G.~A. 2006, Gaia Technical Report GAIA-C8-SP-LEI-AB-006-1

\bibitem[{{Buat} {et~al.}(1999){Buat}, {Donas}, {Milliard}, \& {Xu}}]{buat}
{Buat}, V., {Donas}, J., {Milliard}, B., \& {Xu}, C. 1999, \aap, 352, 371

\bibitem[{Burges(1998)}]{burges98}
Burges, C. J.~C. 1998, Data Mining and Knowledge Discovery, 2, 121

\bibitem[{Chang \& Lin(2001)}]{libsvm}
Chang, C.-C. \& Lin, C.-J. 2001, {LIBSVM}: a library for support vector
  machines, software available at http://www.csie.ntu.edu.tw/~cjlin/libsvm

\bibitem[{{Fioc}(1997)}]{fioc3}
{Fioc}, M. 1997, PhD thesis, Universit{\'e} Paris XI,
  http://www.iap.fr/users/fioc.html

\bibitem[{{Fioc}(1999)}]{fioc6}
{Fioc}, M. 1999, in Astronomical Society of the Pacific Conference Series, Vol.
  192, Spectrophotometric Dating of Stars and Galaxies, ed. I.~{Hubeny},
  S.~{Heap}, \& R.~{Cornett}, 299--+

\bibitem[{{Fioc} \& {Rocca-Volmerange}(1997)}]{fioc2}
{Fioc}, M. \& {Rocca-Volmerange}, B. 1997, \aap, 326, 950

\bibitem[{{Fioc} \& {Rocca-Volmerange}(1999{\natexlab{a}})}]{fioc1}
{Fioc}, M. \& {Rocca-Volmerange}, B. 1999{\natexlab{a}}, \aap, 351, 869

\bibitem[{{Fioc} \& {Rocca-Volmerange}(1999{\natexlab{b}})}]{fioc4}
{Fioc}, M. \& {Rocca-Volmerange}, B. 1999{\natexlab{b}}, \aap, 344, 393

\bibitem[{{Fioc} \& {Rocca-Volmerange}(1999{\natexlab{c}})}]{fioc5}
{Fioc}, M. \& {Rocca-Volmerange}, B. 1999{\natexlab{c}}, arXiv:astro-ph/9912179

\bibitem[{{Fukugita} {et~al.}(1996){Fukugita}, {Ichikawa}, {Gunn}, {Doi},
  {Shimasaku}, \& {Schneider}}]{fukugita}
{Fukugita}, M., {Ichikawa}, T., {Gunn}, J.~E., {et~al.} 1996, \aj, 111, 1748

\bibitem[{{Groenewegen} \& {de Jong}(1993)}]{groenewegen}
{Groenewegen}, M.~A.~T. \& {de Jong}, T. 1993, \aap, 267, 410

\bibitem[{{Le Borgne} \& {Rocca-Volmerange}(2002)}]{le2}
{Le Borgne}, D. \& {Rocca-Volmerange}, B. 2002, \aap, 386, 446

\bibitem[{{Le Borgne} {et~al.}(2004){Le Borgne}, {Rocca-Volmerange},
  {Prugniel}, {Lan{\c c}on}, {Fioc}, \& {Soubiran}}]{le1}
{Le Borgne}, D., {Rocca-Volmerange}, B., {Prugniel}, P., {et~al.} 2004, \aap,
  425, 881

\bibitem[{{Rana} \& {Basu}(1992)}]{rana}
{Rana}, N.~C. \& {Basu}, S. 1992, \aap, 265, 499

\bibitem[{{Rocca-Volmerange} {et~al.}(2007){Rocca-Volmerange}, {de Lapparent},
  {Seymour}, \& {Fioc}}]{rocca07}
{Rocca-Volmerange}, B., {de Lapparent}, V., {Seymour}, N., \& {Fioc}, M. 2007,
  arXiv:0705.2031

\bibitem[{{Rocca-Volmerange} {et~al.}(2004){Rocca-Volmerange}, {Le Borgne}, {De
  Breuck}, {Fioc}, \& {Moy}}]{rocca}
{Rocca-Volmerange}, B., {Le Borgne}, D., {De Breuck}, C., {Fioc}, M., \& {Moy},
  E. 2004, \aap, 415, 931

\bibitem[{Vapnik(1995)}]{vapnik95}
Vapnik, V.~N. 1995, The nature of statistical learning theory (Springer)

\bibitem[{{Williams} {et~al.}(1996){Williams}, {Blacker}, {Dickinson}, {Dixon},
  {Ferguson}, {Fruchter}, {Giavalisco}, {Gilliland}, {Heyer}, {Katsanis},
  {Levay}, {Lucas}, {McElroy}, {Petro}, {Postman}, {Adorf}, \&
  {Hook}}]{williams}
{Williams}, R.~E., {Blacker}, B., {Dickinson}, M., {et~al.} 1996, \aj, 112,
  1335

\bibitem[{{Woosley} \& {Weaver}(1995)}]{woosley}
{Woosley}, S.~E. \& {Weaver}, T.~A. 1995, \apjs, 101, 181

\end{thebibliography}
\end{document}